\documentclass[sigconf, nonacm]{acmart}

\usepackage[normalem]{ulem}

\usepackage{url}
\usepackage{graphicx}
\usepackage{listings}
\usepackage{color}
\usepackage{subfigure}
\usepackage{xspace}
\usepackage{latexsym}
\makeatletter
\newif\if@restonecol
\makeatother

\usepackage[ruled,vlined,linesnumbered]{algorithm2e}
\usepackage{algorithmic}
\usepackage{enumerate}

\usepackage{makecell}

\usepackage{fontawesome}

\usepackage{wrapfig}
\usepackage{setspace}
\usepackage{blindtext}
\usepackage{amsfonts}
\usepackage{multirow}
\usepackage{pifont}
\usepackage{mathrsfs}
\usepackage{mathtools}
\usepackage{relsize}
\usepackage{comment}
\usepackage{footnote}
\usepackage{tikz}
\usetikzlibrary{calc}
\makesavenoteenv{tabular}
\makesavenoteenv{table}
\usepackage{soul}

\usepackage{framed}  
\colorlet{shadecolor}{gray!20}

\usepackage{pifont}

\usepackage{makecell}
\usepackage{multirow}
\usepackage{booktabs}
\usepackage{siunitx}
\usepackage{enumitem}

\definecolor{bostonuniversityred}{rgb}{0.8, 0.0, 0.0}
\definecolor{blush}{rgb}{0.87, 0.36, 0.51}
\definecolor{inchworm}{rgb}{0.7, 0.93, 0.36}
\definecolor{neoncarrot}{rgb}{1.0, 0.64, 0.26}
\definecolor{nadeshikopink}{rgb}{0.96, 0.68, 0.78}
\definecolor{mediumtealblue}{rgb}{0.0, 0.33, 0.71}
\definecolor{lightskyblue}{rgb}{0.53, 0.81, 0.98}
\definecolor{iceberg}{rgb}{0.44, 0.65, 0.82}

\newif{\ifSubmit}
\newif{\ifFinal}
\newif{\ifDraft}
\newif{\ifDraftVLDB}
\Submittrue

\ifSubmit
\newcommand{\alicomment}[1]{}
\newcommand{\yuecomment}[1]{}
\newcommand{\vcomment}[1]{}
\newcommand{\aocomment}[1]{}
\newcommand{\yanc}[1]{}
\newcommand{\lrcomment}[1]{}
\newcommand{\dimcomment}[1]{}
\newcommand{\xlcomment}[1]{}
\else
\newcommand{\alicomment}[1]{\noindent\textcolor{magenta}{\bf Ali: #1}}

\newcommand{\aocomment}[1]{\noindent\textcolor{blue}{\bf Ao: #1}}
\newcommand{\yanc}[1]{\textcolor{brown}{\textbf{Feng: #1}}}
\newcommand{\lrcomment}[1]{\textcolor{teal}{\textbf{Lukas: #1}}}
\newcommand{\dimcomment}[1]{\textcolor{cyan}{\textbf{Dimitris: #1}}}
\newcommand{\xlcomment}[1]{\textcolor{purple}{\textbf{Xiaolong: #1}}}
\newcommand{\vcomment}[1]{\textcolor{yellow}{\textbf{Vasily: #1}}}
\fi

\ifDraft
\newcommand{\addcomment}[1]{\textcolor{red}{#1}}
\newcommand{\diffcomment}[2]{\textcolor{orange}{#1}}
\newcommand{\delcomment}[1]{\textcolor{orange}{\sout{#1}}}
\else
\newcommand{\addcomment}[1]{#1}
\newcommand{\diffcomment}[2]{#1}
\newcommand{\delcomment}[1]{}
\fi

\ifDraftVLDB
\newcommand{\addVLDBcomment}[1]{\textcolor{red}{#1}}
\newcommand{\diffVLDBcomment}[2]{\textcolor{orange}{#1}}
\newcommand{\delVLDBcomment}[1]{\textcolor{orange}{\sout{#1}}}
\else
\newcommand{\addVLDBcomment}[1]{#1}
\newcommand{\diffVLDBcomment}[2]{#1}
\newcommand{\delVLDBcomment}[1]{}
\fi

\newcommand{\proj}{\textsc{InfiniStore}\xspace}  
\newcommand{\prelim}{\textsc{InfiniCache}\xspace}  

\usepackage{booktabs}
\usepackage{siunitx}

\usepackage{zref}
\usepackage{cleveref}
\crefformat{section}{\S#2#1#3}

\microtypecontext{spacing=nonfrench}

\usepackage{balance}
\usepackage[frozencache=true,cachedir=minted-cache]{minted} 
\usepackage{verbments}

\usepackage{natbib}

\newcommand{\kvget}{{\small\texttt{GET}\xspace}}%
\newcommand{\kvput}{{\small\texttt{PUT}\xspace}}%

\everypar{\looseness=-1}

\begin{document}


\title{{\proj}: Elastic Serverless Cloud Storage}

\author{Jingyuan Zhang}
\orcid{0000-0001-9581-1807}
\affiliation{%
  \institution{George Mason University}
  \city{Fairfax}
  \state{VA}
  \postcode{22030}
  \country{USA}
}
\email{jzhang33@gmu.edu}

\author{Ao Wang}
\affiliation{%
  \institution{George Mason Univ., Alibaba Group}
  \city{Fairfax}
  \state{VA}
  \postcode{22030}
  \country{USA}
}
\email{shenlan.wa@alibaba-inc.com}

\author{Xiaolong Ma}
\affiliation{%
  \institution{University of Neveda, Reno}
  \city{Reno}
  \state{NV}
  \country{USA}
}
\email{xiaolongm@nevada.unr.edu}

\author{Benjamin Carver}
\orcid{0000-0002-1574-9300}
\author{Nicholas John Newman}
\affiliation{%
  \institution{George Mason University}
  \city{Fairfax}
  \state{VA}
  \postcode{22030}
  \country{USA}
}
\email{bcarver2@gmu.edu}
\email{nnewman7@gmu.edu}

\author{Ali Anwar}
\orcid{0000-0003-4487-2436}
\affiliation{%
  \institution{University of Minnesota}
  \city{Minneapolis}
  \state{MN}
  \postcode{55455}
  \country{USA}
}
\email{aanwar@umn.edu}

\author{Lukas Rupprecht}
\author{Vasily Tarasov}
\orcid{0000-0003-1424-9977}
\affiliation{%
  \institution{IBM Research}
  \city{San Jose}
  \state{CA}
  \postcode{95120}
  \country{USA}
}
\email{lukas.rupprecht@ibm.com}
\email{vtarasov@us.ibm.com}

\author{Dimitrios Skourtis}
\affiliation{%
  \institution{Redpanda Data}
  \city{San Francisco}
  \state{CA}
  \country{USA}
}
\email{skourtis@soe.ucsc.edu}

\author{Feng Yan}
\orcid{0000-0001-9840-7754}
\affiliation{%
  \institution{University of Houston}
  \city{Houston}
  \state{TX}
  \postcode{77004}
  \country{USA}
}
\email{fyan5@central.uh.edu}

\author{Yue Cheng}
\orcid{0000-0003-1695-4864}
\affiliation{%
  \institution{University of Virginia}
  \city{Charlottesville}
  \state{VA}
  \country{USA}
}
\email{yuecheng@virginia.edu}
\authornote{Corresponding author}

\settopmatter{printfolios=true}


\begin{abstract}
Cloud object storage such as AWS S3 is cost-effective and highly elastic but relatively slow, while high-performance cloud storage such as AWS ElastiCache is expensive and provides limited elasticity. We present a new cloud storage service called ServerlessMemory, which stores data using the memory of serverless functions. ServerlessMemory employs a \diffVLDBcomment{sliding-window-based memory management strategy inspired by the garbage collection mechanisms used in the programming language}{time-window-based data placement strategy} to effectively segregate hot/cold data and provides \diffcomment{fine-grained}{high} elasticity, \addcomment{good} performance, and a pay-per-access cost model with extremely low cost.
%

We then design and implement {\proj}, 
a persistent and elastic cloud storage system, which seamlessly couples the function-based ServerlessMemory layer with a persistent, inexpensive cloud object store layer. {\proj} enables durability despite function failures using a fast parallel recovery scheme built on the auto-scaling functionality of a FaaS (Function-as-a-Service) platform.
We evaluate {\proj} extensively using both microbenchmarking and two real-world applications. Results show that {\proj} has more performance benefits for objects larger than 10 MB compared to AWS ElastiCache and Anna, and {\proj} achieves $26.25\%$ and $97.24\%$ tenant-side cost reduction compared to {\prelim} 
and ElastiCache, respectively. 
\end{abstract}

\maketitle

\vspace{-6pt}
\section{Introduction}
\label{sec:intro}

Public clouds free tenants from the tedious tasks of IT infrastructure planning and maintenance and allow tenants to focus on application development. 
These offerings are driving the adoption of public clouds for hosting massive-scale, data- and compute-intensive applications, such as Internet-scale web applications~\cite{haystack_osdi10, docker_fast18, dockerhub, faast_socc21}.


%
Although cloud providers can simplify the allocation and scaling of compute resources, there is an excessively wide range of cloud \emph{storage} services with various persistence, performance, pricing, and capacity characteristics to choose from. This choice complicates resource management and application deployment. For example, AWS ElastiCache~\cite{aws_ec} is an AWS-managed memory cache service, but ElastiCache does not provide data persistence by default. AWS S3~\cite{aws_s3} is a ubiquitous object store service, which offers data durability and persistence, but S3 is much slower than ElastiCache. 

This choice 
is further complicated by the varying memory and storage requirements of heterogeneous cloud workloads. For example, a production object store workload for accessing Docker container images~\cite{docker_fast18} exhibits a frequently changing working set size (WSS) having a wide range of object sizes and strong temporal reuse patterns---most object reuse happens within one or two hours of the object's previous access.
In addition, emerging serverless function applications require large, short-term, elastic storage capacity that scales based on WSS and object sizes~\cite{faast_socc21}.
Similar dynamic I/O patterns are observed in enterprise network file system workloads~\cite{nfs_wl_atc08}, 
big data analytics workloads~\cite{alibaba_dag_socc19,  pocket_osdi18}, and 
data warehouse workloads~\cite{snowflake_nsdi20}, among others~\cite{aws_aurora_sigmod18, starling_sigmod20}. 
%
%

We argue that today's clouds are missing an elastic, performant, and cost-effective cloud storage solution that can fulfill the heterogeneous and dynamic storage requirements of a wide variety of applications.
Durable cloud object store solutions such as AWS S3~\cite{aws_s3} and Google Cloud Storage~\cite{gcs} are inexpensive but cannot provide memory-store-level performance. Faster storage/caches such as AWS FSx~\cite{aws_fsx} and ElastiCache~\cite{aws_ec} offer high-bandwidth and low-latency data access, but these solutions are expensive and lack the capability to automatically and rapidly grow and shrink storage capacity in response to changing application demands. 
The Function-as-a-Service (FaaS) model is well suited to fill this gap.
%
%
FaaS applications are structured as a collection of \emph{functions} managed by the service provider in terms of resource scaling and management. 
The combination of instant \diffcomment{scaling}{scale-out/scale-in}, fast access, and pay-per-use pricing 
makes FaaS platforms an appealing foundation for an elastic, performant, and cost-effective storage service.

{\prelim} is a distributed memory caching system that exploits the properties of FaaS~\cite{infinicache_fast20}. {\prelim} uses many serverless functions whose function-memory is used collectively for object storage with {\kvput}/{\kvget} APIs for accessing the objects stored. Serverless functions are transient and may be reclaimed after a short period by the FaaS provider. Therefore, {\prelim} implements a primary/backup replication protocol to increase data durability at the serverless function-memory level.


While {\prelim} has demonstrated the feasibility of using serverless functions for data caching, {\prelim} has several limitations. 
(1)~The serverless-function-based cluster deployed in {\prelim} is fixed, and therefore, lacks elasticity: {\prelim} randomly assigns old and new data objects to function instances; if were to be scaled out, this data mapping strategy may lead to excessive data migration, which we will show later.
(2)~{\prelim} provides best-effort data durability via erasure coding and replication at the serverless function level; however, the FaaS provider may reclaim a function instance and its memory at any time, which causes cache misses and impacts application performance.
(3)~Data cached in {\prelim} is replicated twice, which doubles memory resources and doubles cost.


\diffVLDBcomment{In this paper, we introduce a new storage service named
\textit{ServerlessMemory}.
ServerlessMemory uses the collective function-memory as a storage medium to construct a continuous memory space. When allocating memory, new functions are invoked and function-memory are added to memory space. Inspired by the mark-compact garbage collection (GC) algorithms in programming languages~\cite{GC}, we apply a fully-automatic, sliding-window-based function-memory management scheme to separate live/unused (hot/cold) data
and leverage FaaS provider's function reclaiming mechanism for garbage collection. Together, the ServerlessMemory achieves high storage elasticity at function granularity. 
Differentiated with a cache, which would typically require an expensive offline tuning process to construct the miss ratio curve to find the optimal cache size for a certain workload, the ServerlessMemory service can automatically capture the application’s working set.}{} 
\addVLDBcomment{\emph{To the best of our knowledge, the ServerlessMemory is the first cloud service that leverages the desirable FaaS properties to achieve fine-grained elasticity in disaggregated memory management.}}  

\diffVLDBcomment{We build {\proj}, an elastic, fault-tolerant cloud storage 
on top of the ServerlessMemory service. 
{\proj} has two layers: a \textit{ServerlessMemory} layer that exposes durable serverless function-memory to serve application I/Os,
and an inexpensive \textit{object store} layer that uses a cloud object store for data persistence. We implement durability via fast \emph{parallel recovery} and lightweight \textit{insertion logs}. When a function instance is reclaimed unexpectedly, {\proj} launches a group of pre-selected peer (recovery function) instances for parallel data recovery. Each recovery function instance replays a portion of its assigned insertion log and downloads lost data from {\proj}'s object store layer.}{}

This paper makes the following contributions: 
\vspace{-4pt}
\begin{itemize}[noitemsep,leftmargin=*]
    \item We introduce a new ServerlessMemory cloud service that is elastic and pay-per-access at the memory storage level. \addVLDBcomment{The ServerlessMemory is the \emph{first} cloud service that exploits the FaaS properties to automatically capture the working set of a stateful data-intensive application.
    }
    \item We {design and implement} {\proj}, an elastic,  cost-effective, high-performance, \diffcomment{and fault-tolerant}{, and automated} cloud storage system that combines the ServerlessMemory layer with a persistent but inexpensive object store layer. 
    \item We perform extensive evaluations using \diffVLDBcomment{YCSB microbenchmark stress testing}{microbenchmarks} and \addcomment{two practical applications}: an IBM container registry workload and an Azure Functions blobs workload. 
\end{itemize}

Experimental results show that {\proj} represents a novel performance-\$cost tradeoff in today's cloud storage landscape. It is worth noting that {\proj} is a memory storage while reducing cost by $26.25\%$ compared to {\prelim}, $97.24\%$ compared to AWS ElastiCache, and offering better performance for large object requests than Anna~\cite{anna}, AWS S3, and FSx.
{\proj} is pay-per-access, which means it incurs a cost proportional to the number of {\kvget} and {\kvput} requests it serves, with a small cost overhead of $26.00\%$. 
\vspace{-10pt}
\section{Motivation}
\label{sec:moti}

\begin{figure*}[t]
\begin{center}
\subfigure[Working set size timeline.] {
\includegraphics[width=.24\textwidth]{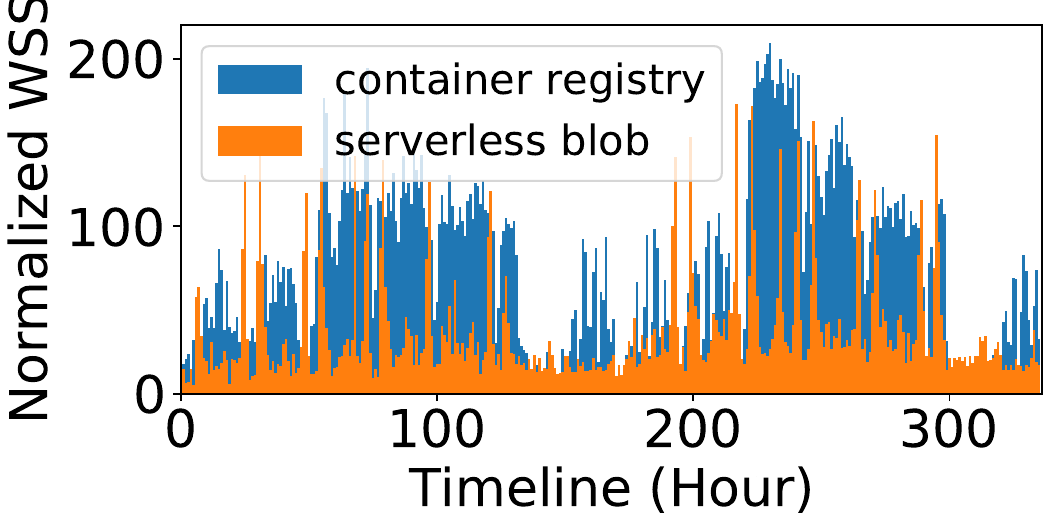}
\label{fig:workload_wss}
}
\hspace{-7pt}
\subfigure[Aggregate I/O throughput timeline.] {
\includegraphics[width=.24\textwidth]{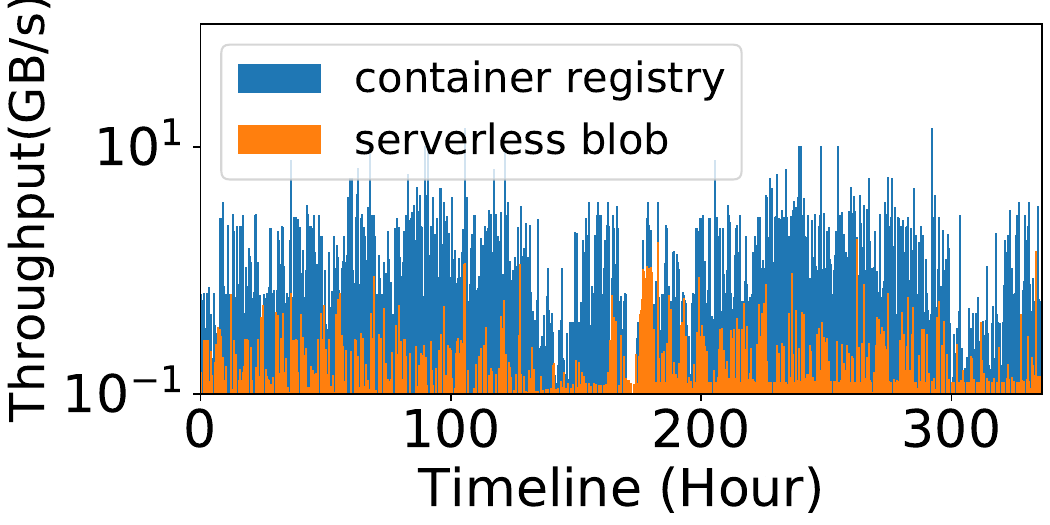}
\label{fig:workload_thpt}
}
\hspace{-7pt}
\subfigure[Reuse interval distribution.] {
\includegraphics[width=.235\textwidth]{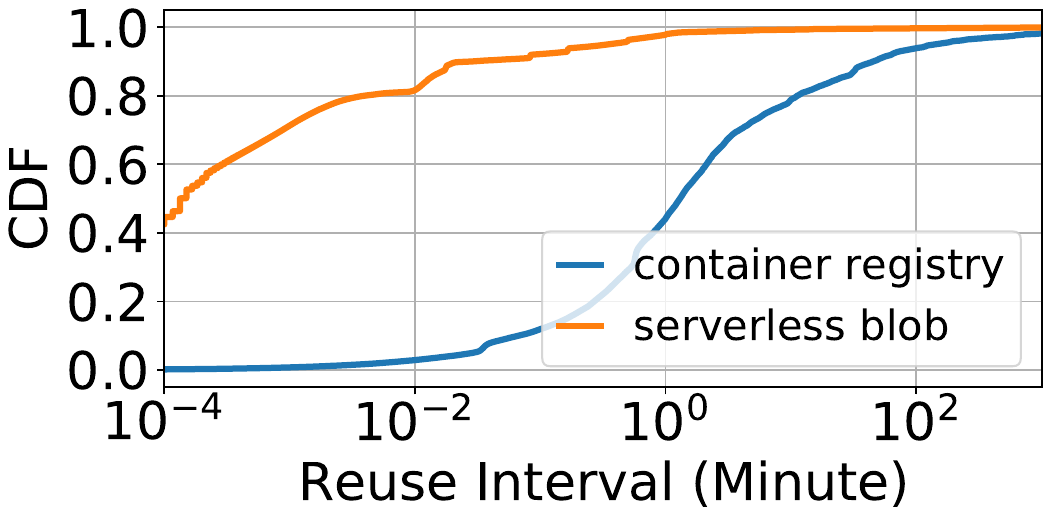}
\label{fig:workload_reuse_cdf}
}
\hspace{-7pt}
\subfigure[IAT CoV distribution ($\# reuse\geq10$).] {
\includegraphics[width=.235\textwidth]{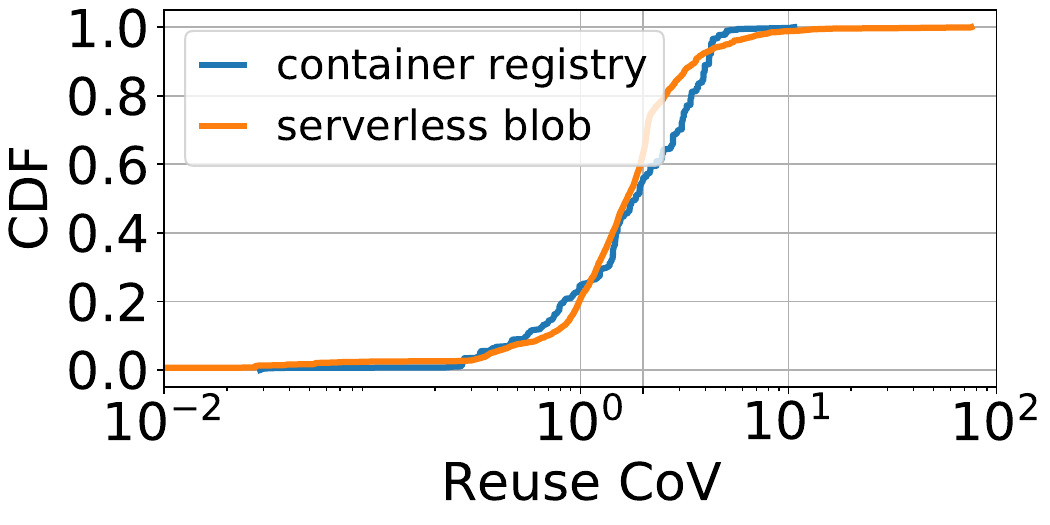}
\label{fig:workload_reuse_cov}
}
\vspace{-12pt}
\caption{The container registry and serverless blob workload characteristics: dynamicity (a, b) and temporal (c, d) behaviors.}
\label{fig:moti_wl}
\end{center}
\vspace{-10pt}
\end{figure*}

Data-intensive applications have dynamic and heterogeneous workloads that benefit from storage elasticity. 
This section performs a detailed workload analysis on two representative storage workloads: a cloud object store workload of IBM container registry~\cite{docker_fast18, bolt_cloud19} and a serverless application workload of Azure Functions blob accesses~\cite{faast_socc21}. To understand the dynamic characteristics of these workloads,
we will focus along two dimensions: 
(1) working set size (WSS) and throughput; and (2) temporal access patterns.
WSS here means the aggregate footprint of the data accessed, re-accessed, and modified by the application in a given time interval. 

\vspace{-6pt}
\subsection{Dynamic WSS and Throughput}
\label{subsec:moti-wss-throughput}

We begin our workload analysis by asking these questions:
\begin{enumerate}[noitemsep,leftmargin=*]
    \item \emph{Does WSS/aggregate throughput dynamically change?}
    \item \emph{If so, what is the magnitude of this change?}
\end{enumerate}

The answers to these questions will help us understand the elasticity requirement for {\proj}. 

Both workloads exhibit a highly variable WSS with a maximum size over $209\times$ and $173\times$ larger than the minimum size for the container image workload and serverless application workload, respectively (Figure~\ref{fig:workload_wss}).
Additionally, the WSS is shifting every minute.
Such significant temporal WSS variance suggests that a high degree of elasticity at the storage level is required to be able to serve the objects in the working set efficiently without excessive storage overprovisioning.

The container image workload also demonstrates bursty I/O
characteristics. Figure~\ref{fig:workload_thpt} shows that the aggregate 
throughput spikes to 15 GB/s with an average of 1 GB/s. This throughput variability
poses a significant challenge for tenants trying to provision storage resources that meet the application's performance requirements.

\noindent\textbf{Implication 1:} \emph{The WSS variance requires instantaneous provisioning of large amounts of storage resources to satisfy the high throughput and low latency requirements.}

\vspace{-6pt}
\subsection{Temporal Access Pattern}
\label{subsec:moti-temporal}
We study the temporal access patterns by asking:

\begin{enumerate}[noitemsep,leftmargin=*]
    \item \emph{What is the 
    time interval between two successive accesses to the same data object?}
    \item \emph{What is the request inter-arrival time (IAT) pattern for reused objects?} 
\end{enumerate}

The answers to these questions will guide the design of {\proj}'s elastic data placement strategy, i.e., the mapping between data objects and serverless functions.

We first study the temporal patterns by measuring the time interval between two successive accesses of the same data object
(the temporal reuse interval).  
Figure~\ref{fig:workload_reuse_cdf} shows that for the first 50 hours of the container image workload, approximately $80\%$ and $94\%$ of all reuses happen within 10 minutes and 100 minutes, respectively. 
The serverless application workload shows a similar reuse pattern but with much shorter reuse intervals: approximately $98\%$ of requests revisit the same object within one minute.
These patterns suggest that much of the data accessed is very likely to be reused within a short time interval.

We next study the IAT pattern by quantifying the coefficient of variation (CoV) of reused data objects.
A CoV of 1 suggests that a re-access may arrive at an arbitrary time, where the IAT follows a Poisson distribution. A CoV $>1$ indicates a more bursty arrival pattern than a Poisson distribution.
We filter out objects with less than 10 reuses, which excludes 16\% and 2\% of the requests from the container image and serverless blob workloads, respectively.  Figure~\ref{fig:workload_reuse_cov} shows that 
about $80\%$ of reused objects have a CoV greater than 1, indicating that both workloads are very bursty.

\noindent\textbf{Implication 2:} \emph{A multi-layer storage system with a fast elastic layer and a slower persistence layer would be beneficial. Most data become cold quickly and can be discarded from the faster storage layer 
if all data are durably stored in a persistence layer. This also suggests that \diffVLDBcomment{hot data must be tracked and hot/cold data segregation would be beneficial in this context.}{a FIFO-like (first in, first out) data placement policy that segregates old (cold) and new (hot) data objects would work well in this context.}}

\vspace{-6pt}
\section{Why Use FaaS for Data Storage?}\label{sec:faas}
\subsection{FaaS Properties}
\label{subsec:serverless}
%

The above implications motivate us to rethink the design and implementation of cloud storage systems. The main question we seek to answer is: \emph{How can a cloud storage system be designed to offer elasticity for dynamically changing workloads while simultaneously providing high performance and high durability at low cost?} 
%
In this section, we first introduce the unique properties of FaaS and explain how storage services that are built on cloud functions can take advantage of these properties.
We then discuss {\prelim} and its limitations.


\noindent\textbf{Quick Startup and Pay-per-Use.}
Unlike VMs, which can take minutes to launch, thousands of cloud function instances can be provisioned in a fraction of a second, without advance notice, via an HTTP API. Storage services can use fast and massive instance provisioning to effectively handle bursty and dynamically-changing storage access patterns.
%
%
More importantly, FaaS providers \cite{lambda, google_func, ibm_func} charge tenants on a fine-grained per-use basis. For example, AWS Lambda bills on a per-invocation basis ($\$0.02$ per 1 million invocations) and charges for a  CPU+memory bundle at a rate of $\$0.0000166667$ per second (rounding up to the nearest millisecond) for each GB of bundled memory.
This means accessing objects stored in the memory of function instances will be billed on a per-access basis.
%
%

\noindent\textbf{Exploiting Opportunistic and Elastic Function Memory for Data Storage.}
Recent studies~\cite{peeking_atc18, infinicache_fast20} report that FaaS providers support limited, short-term caching of function instances by keeping the instance states ``warm'' in memory to mitigate cold start penalties~\cite{cold_start_war}.
Idle function instances that are not invoked after a period of time can be reclaimed by the provider. This period of time varies, ranging from tens of minutes to hours for AWS Lambda. 
A later invocation to the same instance can extend the function's lifetime. 
The memory of a cluster of function instances can be aggregated to enable opportunistic storage services.
Furthermore, fast function startup enables low-latency resource scaling when more functions need to be provisioned,
which provides an elastic storage foundation for highly dynamic workloads.
%

\vspace{-6pt}
\subsection{\prelim}
\label{subsec:infinicache}

\begin{figure}[t]
\begin{center}
\vspace{-5pt}
\includegraphics[width=0.42\textwidth]{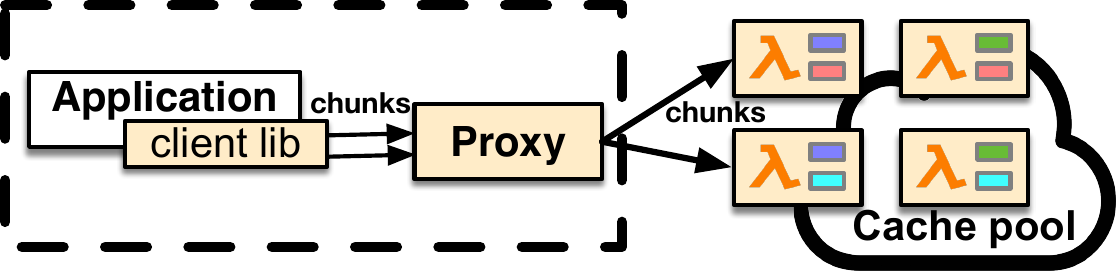}
\vspace{-15pt}
\caption{{\prelim} architecture.}
\label{fig:archIC}
\end{center}
\vspace{-15pt}
\end{figure}

\begin{table}[h]
\begin{center}
\vspace{-10pt}
\scalebox{0.8}{
\begin{tabular}{ll}
{\bf Terminology} & {\bf Definition} \\
\Xhline{2\arrayrulewidth}
{\bf Chunk} & {A partition of an erasure-coded data object} \\
{\bf Function} & {A deployment of AWS Lambda} \\
{\bf Instance} & {A Lambda instance w/ bundled compute-memory resources} \\
{\bf Function-memory} & {The memory resource in a function instance} \\
\end{tabular}
}
\caption{\addVLDBcomment{Summary of terminologies used in {\prelim}.}}
\label{tbl:terms}
\vspace{-20pt}
\end{center}
\end{table}

\noindent{\prelim}~\cite{infinicache_fast20} builds on the above insights by exploiting the collective memory of serverless functions to cache data objects. Figure~\ref{fig:archIC} shows the architecture of {\prelim}. {\prelim} exposes {\small\texttt{\kvget}}/{\small\texttt{\kvput}} API via a client library. 
\diffVLDBcomment{To {\small\texttt{\kvput}} an object, the client library sends erasure-coded \emph{\textbf{chunks}} of the object to a proxy. The proxy randomly maps chunks to Lambda \emph{\textbf{function deployments}} in the cache pool and streams data to invoked  \emph{\textbf{function instances}} for chunk storage.
The chunk-function mapping is stored in the proxy for serving {\small\texttt{\kvget}} requests.}{
Applications send {\small\texttt{\kvget}} requests for an object to a proxy. The proxy locates Lambda \emph{\textbf{function}} deployments in the cache pool each storing an erasure coded chunk of the object.
The proxy then invokes these function \emph{\textbf{instances}} to access the object chunks stored in their \emph{\textbf{function-memory}}.}
\addVLDBcomment{The terms are summarized in Table~\ref{tbl:terms}.}

{\prelim} periodically invokes all functions in the cache pool to keep idle function instances from being reclaimed by AWS. {\prelim} uses two fault tolerance techniques, \diffVLDBcomment{primary/backup delta-replication and erasure coding, to maximize data durability at the cache level. The delta-replication technique replicates all chunks twice between two instances of the same function deployment. The replication provides instantaneous failover if one of the two replica instances is failed. Erasure coding complements replication by providing object-level fault tolerance
if no more than a configurable fraction of an object's chunks are lost.}{
erasure coding and primary/backup delta-replication, to maximize data durability at the cache level. 
Objects are stored as erasure-coded chunks. Therefore, {\prelim} can recover from a data loss if no more than a configurable fraction of data chunks is lost (due to instance reclamation). The delta-replication technique replicates all data chunks twice across two instances of the same function, resulting in a primary instance and a backup instance. This replication allows instantaneous failover when the primary instance is reclaimed.}

\noindent\textbf{{\prelim}'s Limitations:}
\label{para:ic_limitation}
\vspace{-2pt}
\begin{itemize}[noitemsep,leftmargin=*]
\item Lack of elasticity: {\prelim} maintains a static, fixed cache pool of functions whose collective function-memory is typically larger than the active WSS of the workload. This strategy may cause massive data movement if the cache pool were scaling out and data rebalancing is required.
\item Mixed hot/cold object placement \addVLDBcomment{and no hot data tracking}: {\prelim} arbitrarily assigns new data objects to functions in the cache regardless of whether the data will remain hot or become cold. Hot data \diffVLDBcomment{must be tracked and migrated}{will have to be migrated} if a function were selected to be removed from the cache on scaling-in events.
\item Durability is not guaranteed:
Function instances reclamation causes {\prelim} to suffer data loss at the memory cache level. {\proj} allows no data loss as a memory storage.
\end{itemize}
\vspace{-6pt}

\vspace{-4pt}
\section{Serverless Memory}
\label{sec:svrlessmem}

We propose a new cloud storage service
called \emph{ServerlessMemory}, which combines the memory of a cluster of serverless function instances and exposes this memory to applications as fast, elastic, inexpensive, and pay-per-access cloud storage. To provide durability, {\proj} combines ServerlessMemory with a persistent backing object store.
In this section, we provide an overview of ServerlessMemory.

In ServerlessMemory, cloud function instances and their supporting function-memories are automatically allocated when the tenant application calls the {\small\texttt{PUT}} API. This abstraction is
analogous to programming languages that provide automatic memory management, e.g., Java. 
%
Data objects are inserted into the memory of function instances and accessed by re-invoking functions. 

When an application’s storage demand surges, more ServerlessMemory functions can be invoked. The function-memory of the new function instances joins the distributed memory pool instantly without requiring any manual effort for launching and scaling servers that are otherwise hosted by cloud VMs. 
A function $F$'s function-memory that stores data object $O$
is only billed when $O$ is accessed by a {\kvget} or {\kvput} operation of $F$, i.e., when an instance of $F$ is executing. This pay-per-access property is naturally utilized by ServerlessMemory.
%
%
\addVLDBcomment{Hot data are tracked by the ServerlessMemory and compacted into a collection of active function instances.}
If $F$ \addVLDBcomment{is not in the collection and $F$'s} function-memory holds cold data, $F$ is rarely invoked, and the storage for the data objects in $F$'s function-memory is rarely billed.
No explicit garbage collection is required for ServerlessMemory as storage is implicitly garbage collected by the serverless provider 
when the instance of $F$ becomes inactive for a prolonged period and is reclaimed. 
\vspace{-6pt}
\section{{\proj} Design}
\label{sec:design}

{\proj} is a co-designed cloud storage system that tightly couples a serverless, function-based ServerlessMemory store (SMS) layer and a persistent cloud object store (COS) layer. All data is stored in SMS for fast access and copied to COS for durability.
SMS is highly configurable and implements an adaptive, sliding-window-based data management mechanism to hold the current working set of a workload. 
In this section, we present the design of {\proj} and its SMS and COS components.

\vspace{-6pt}
\subsection{SMS Design Challenges}
\label{subsec:design_challenges}

%
ServerlessMemory poses two design challenges: elasticity-optimized data placement and data durability.

\noindent\textbf{Data Placement.} The strategy that assigns data objects to functions is critical as it impacts the elasticity and cost-effectiveness of the storage system. Randomly placing data in function instances using conventional (serverful) strategies, such as consistent hashing~\cite{ch_stoc97}, may lead to unnecessary expenses. 
As cold data would be co-located with hot/new data in the same function instance,
%
%
function warmups 
would be needed for the entire ServerlessMemory instance pool, which increases monetary cost.
Hence, we need to design an elasticity-optimized 
data placement strategy for ServerlessMemory to minimize function-warmup overhead.
%
%
%

\noindent\textbf{Data Durability in Functions.} 
As mentioned in \cref{para:ic_limitation}, \prelim does not guarantee durability and best-effort data recovery impacts cost-effectiveness.
ServerlessMemory must handle data durability issues 
transparently without noticeably impacting the application's performance and the monetary cost. 
To effectively meet this objective, we combine SMS with an inexpensive COS.

\begin{figure}[t]
\begin{center}
\includegraphics[width=0.44\textwidth]{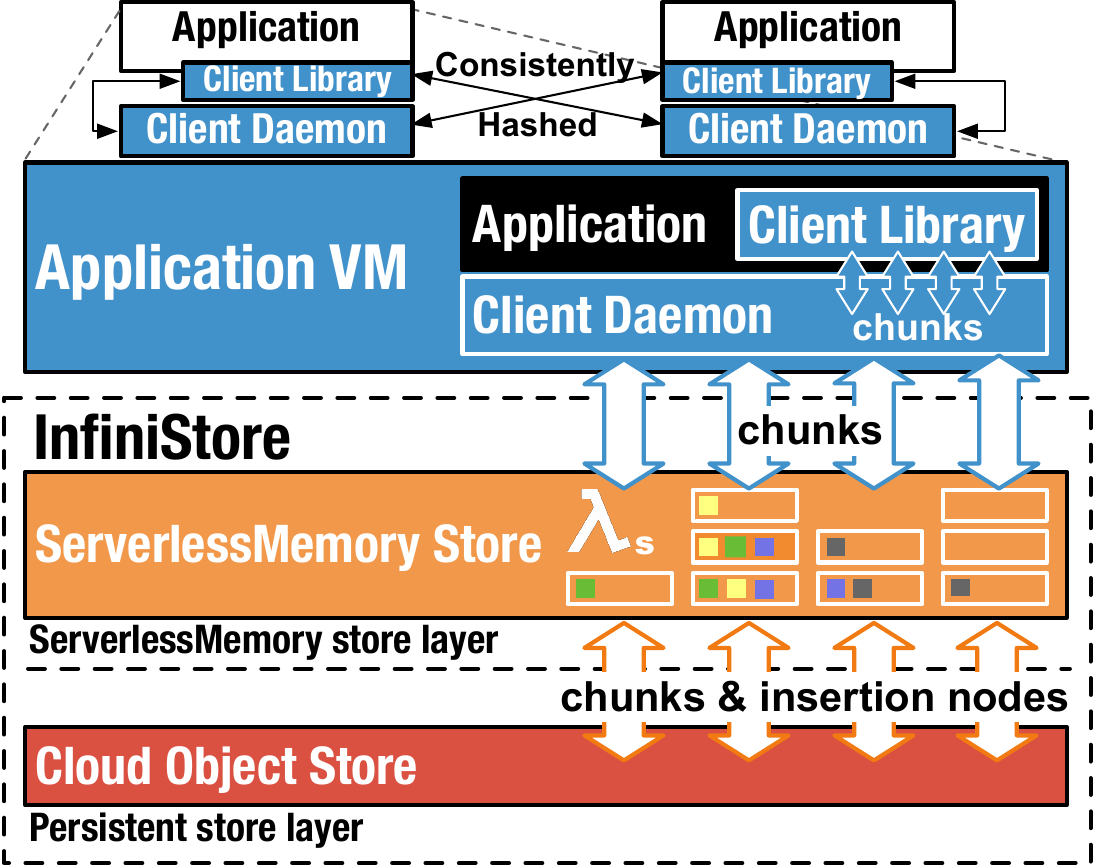}
\vspace{-8pt}
\caption{{\proj} architecture.
}
\label{fig:arch}
\vspace{-18pt}
\end{center}
\end{figure}

\vspace{-6pt}
\subsection{Design Overview}
\label{subsec:design_overview}

Next, we present {\proj}'s design.
We use AWS Lambda as an example to illustrate the design of {\proj}.
In the remainder of this section, all instances 
are assumed to be Lambda function instances.
It is worth noting that the design of {\proj} is generic and can be easily ported to other cloud platforms. 
{\proj} consists of four components: a {\proj} client library, a client daemon, 
an SMS layer, and a persistent COS layer, see Figure~\ref{fig:arch}.

Applications interact with {\proj} via a \textbf{client library} that communicates with the client daemon managing the SMS. The library exposes 
{\small\texttt{GET(key)}}/{\small\texttt{PUT(key, value)}} as read/write APIs
to the application and is responsible for (1)~transparently encoding and decoding data objects with Reed-Solomon erasure coding (EC), and (2)~load balancing incoming application requests across a distributed set of client daemons. 
{\proj} offers strong consistency. The {\kvput} API uses versioning provided by the client daemon to support object updates. Versions of objects are read-only after {\kvput}s return.

Co-located with the application, a \textbf{client daemon} stores data objects' metadata with the versioning information in an in-memory data structure called metadata table, orchestrates the function instances of SMS, and serves as a rendezvous point for streaming EC-encoded object chunks between the client library and SMS. The metadata table can be persisted to the local disk for fault tolerance.
In the distributed application setting, we adopt a multi-VM deployment in which a client daemon is deployed on each VM that hosts the application and manages a separate SMS with shared access among application clients.

\textbf{SMS} consists of \diffcomment{a collection of function instances}
{a large number of function instances}. \addVLDBcomment{Unlike {\prelim}, in SMS, each function instance does not have a peer replica.} 
The scaling of SMS is driven by the workload's working set. SMS manages the active data objects in \diffcomment{function-memory}{the memory of the function instances} and serves all requests sent from the client daemon (\cref{subsec:sms}, \cref{subsec:buffering}).

\textbf{COS} forms the persistence layer of {\proj}. COS stores all data objects and critical metadata (insertion logs) for SMS data recovery (\cref{subsec:recovery}).
To persistently store an object, {\proj}'s client library first determines the destination client daemon (and therefore its corresponding SMS) using consistent hashing. The client library then streams EC-encoded chunks to SMS via the daemon. 
Without compromising the durability and strong consistency, we design a persistent buffer (\cref{subsec:persistent_buffer}) to allow the client application to receive a response once all chunks are successfully inserted into SMS but before all chunks are fully persisted to COS. The function instances in SMS will not return until chunks are fully written to COS.

\vspace{-6pt}
\subsection{ServerlessMemory Store Management}
\label{subsec:sms}

{\proj}'s SMS management design is based primarily on the workload characteristics described in \cref{sec:moti} and the requirement to provide elasticity, performance, 
and durability (\cref{subsec:infinicache}). 
Instead of a static approach, {\proj}'s client daemon uses a novel, highly adaptive, sliding-window-based SMS management mechanism \addVLDBcomment{inspired by the garbage collector designs used in programming languages}. 
The ServerlessMemory is regarded as a continuous memory space of functions, where each function is identified by a global unique $ID_{\lambda}$. FaaS platforms offer virtually infinite memory capacity and new memory can be allocated by simply invoking more functions. In a garbage-collection-based (GC-based) programming language, a GC procedure is invoked once every fixed time interval to release the allocated memory that is no longer referenced. In {\proj} settings, ``no longer referenced'' data is cold data that has not been accessed for a specified time period $H$. Cold data's memory can be released by changing the function management policy (e.g., by stopping invoking corresponding functions and leaving the instances to be eventually reclaimed by the FaaS provider). The client daemon organizes the memory space at the function granularity. 
New data is always appended to functions newly added to memory space. The daemon adds new functions when needed (e.g., out of function-memory, \cref{subsec:puts}). Data re-accessed within $H$ is marked and compacted to newly added functions, too (\cref{subsec:gets}). These functions added during the same GC interval construct a new \emph{GC-bucket}. With compaction, a GC-bucket contains only cold data after $H$ and the memory-functions within is released by the GC.

\begin{figure}[t]
\begin{center}
\vspace{-3pt}
\includegraphics[width=0.48\textwidth]{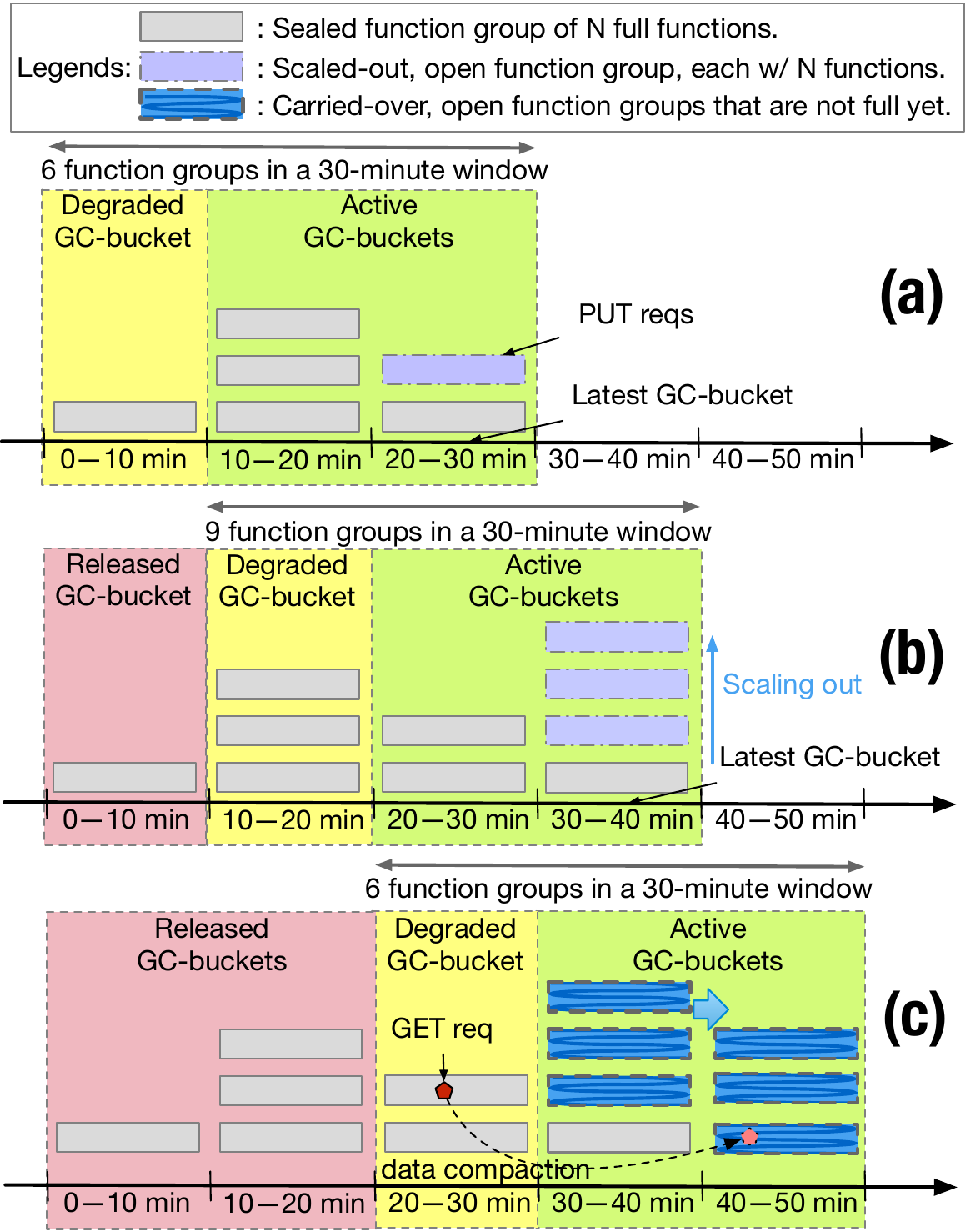}
\vspace{-20pt}
\caption{A sliding-window example of GC execution. 
The example shows a configuration in which \diffcomment{GC executes}{a new I-bucket is created} every 10 minutes, and the active ($M$) and degraded ($N$) window are configured as 20 and 10 minutes respectively. $H$ in this case is 30 minutes.
Functions are shown as FGs.
(a) SMS has been running for 20+ minutes, with its latest GC-bucket in the third time slot.
(b) Time runs into the {\small\texttt{30-40 min}} slot. The \diffcomment{GC}{client daemon} releases the oldest GC-bucket. More functions are added to the latest GC-bucket due to scaling out.
(c) At {\small\texttt{40-50 min}}, The \diffcomment{GC}{client daemon} releases one more GC-bucket. Functions not fully populated keep open in the new GC-bucket.
}
\label{fig:time_mgmt}
\vspace{-15pt}
\end{center}
\end{figure}

For a smoother function management policy (FMP) changing across the function lifespan, the client daemon divides functions added within $H$ into active and degraded GC-buckets, each lasting for $M$ and $N$ GC intervals, respectively\footnote{$M$ and $N$ are configurable. Empirically, $M$ and $N$ can be configured based on the $mean$ and $mean+stdev$ of objects' reuse interval, respectively.}. Figure~\ref{fig:time_mgmt} uses a sliding-window example to illustrate how {\proj}'s GC works. For functions added to memory within $M$ GC intervals in active GC-buckets, we apply an active FMP that extends instance lifespan using a no-op heartbeat message (warmup) sent periodically by the daemon. On executing the GC procedure,
(1)~the oldest active GC-bucket created $M$ intervals ago becomes \emph{degraded}, which is then applied a degraded FMP with a reduced warmup frequency, 
(2)~the oldest degraded GC-bucket created $H$ intervals ago becomes \emph{released} and all functions in released GC-bucket are immediately removed from memory space. Any function in a degraded GC-bucket will be removed 
if failures are detected (\cref{subsec:failure-detection}).

\noindent\textbf{Function Group.}
Within a GC-bucket, the client daemon manages functions by \emph{function groups} (FGs). An FG is a logical scaling unit that consists of $N$ functions, where $N$ is determined by the number of chunks of an object. Note that this abstraction is essential to support objects with various numbers of chunks. The client daemon scales out the latest GC-bucket at FG granularity. 

For clarification, the client daemon serves {\kvput}/{\kvget} requests at the chunk granularity. All chunks of an object are served in parallel.

\vspace{-6pt}
\subsubsection{Serving {\kvput}s}
\label{subsec:puts}

As shown in Figure~\ref{fig:time_mgmt}(a), {\kvput}s are served by the latest GC-bucket. FGs in the current GC-bucket that actively serve {\kvput}s are referred to as \emph{open} FGs. When the workload surges, e.g., the WSS or the number of concurrent {\kvput} requests increases, the daemon scales out the latest GC-bucket by launching more FGs (Figure~\ref{fig:time_mgmt}(b)).  
The client daemon keeps track of the memory consumption of each function in the latest GC-bucket by reconciling the memory statistics piggybacked
on the response of each function invocation request. When the memory consumption of an FG in the latest GC-bucket exceeds a predefined {\small\texttt{HARDCAP}} threshold\footnote{\addVLDBcomment{The {\small\texttt{HARDCAP}} is defined by excluding the Lambda function's program and runtime overhead (around 100~MB) and a fraction of the total function-memory reserved for data recovery (\cref{subsec:parallel_recovery}).}}, the daemon starts the scale-out process and all functions in that FG are sealed (read-only). 
Note that the degraded GC-buckets along with the functions therein are also \emph{sealed}. 

\begin{figure}[t]
\begin{center}
\begin{pyglist}[fontsize=\footnotesize, numbers=left, numbersep=2pt, language=go]
func PlaceChunk(chunk *Chunk):
  // Initialize a func pointer using the object's chunk ID.
  funcPtr := chunk.ID
  // Ensure at least funcPtr open functions are available.
  functions := GetOpenFuncs(funcPtr)
  for:
    if funcPtr >= len(functions): // Needs to scale out?
      // Scaling out by ensuring funcPtr open functions available.
      functions, funcPtr = GetOpenFuncs(funcPtr)
    else if !TestAndPlace(functions[funcPtr], chunk):
      funcPtr += chunk.FGSize // Increment func pointer by one FG.
    else: return
\end{pyglist}
\vspace{-10pt}
\caption{
Chunk placement algorithm. 
}
\label{fig:placement}
\vspace{-18pt}
\end{center}
\end{figure}

\noindent\textbf{Data Placement under {\kvput}s.}
{\proj}'s SMS management employs a simple and highly-efficient data placement algorithm, see Figure~\ref{fig:placement}.
To {\kvput} a new object into SMS, the client daemon calls {\small\texttt{PlaceChunk()}} in parallel for each of the $N$ chunks ({\small\texttt{chunk.FGSize}}) in order to determine in which FG the chunk should be stored. 
\diffVLDBcomment{{\small\texttt{PlaceChunk()}} tests FGs using a function pointer ({\small\texttt{funcPtr}}). Starting from the ${ID_{chunk}}^{th}$ function (line 3), the algorithm first ensures that there are at least {\small\texttt{funcPtr}} functions open for placement (line 5). The algorithm then tries to place the chunk in the function identified by {\small\texttt{funcPtr}} using {\small\texttt{TestAndPlace()}} atomically (line 10, first iteration). If failed, the algorithm advances {\small\texttt{funcPtr}} by $N$ (line 11) to ensure that at least {\small\texttt{funcPtr}} functions open (line 9 in later iterations), auto-scales if needed, and probes the next FG (line 10) until it succeeds. The {\small\texttt{funcPtr}} is advanced by FGs to ensure that}{
{\small\texttt{PlaceChunk()}} first identifies an FG that has enough functions to serve $O$ chunks. The algorithm then checks whether the chunk can be mapped to a unique function within this FG.} {\proj} never places any two chunks of an object on the same function for two reasons: 
(1)~to minimize the chances of multiple chunks becoming unavailable due to a single function reclamation;
and (2)~to parallelize the I/Os across all function instances that will store the object chunks. 
To balance the load and minimize network contention, {\small\texttt{TestAndPlace()}} 
validates (1) if the memory {\small\texttt{HARDCAP}} of the function is reached, 
and (2) if this function's request queues (\cref{subsec:queue}) are full; 
Otherwise, the function serves the {\small\texttt{PUT}} request.
{\small\texttt{PlaceChunk()}} uses a greedy policy for choosing an available FG within the latest GC-bucket. Specifically, the algorithm always tries to use the oldest open FGs for inserting new data. This design is because older FGs holding relatively cold data are likely to be first to reach their memory {\small\texttt{HARDCAP}} and thus are likely to be sealed earlier than newer FGs. All the \emph{open} FGs are carried over to the new GC-bucket during GC, as depicted in Figure~\ref{fig:time_mgmt}(c).

\noindent\textbf{Auto-Scaling under {\kvput}s.}
\proj support customized auto-scaling policies in {\small\texttt{GetOpenFuncs()}} to handle {\kvput} spikes (e.g., one can implement a more aggressive policy that doubles the number of functions each time an auto-scaling is triggered~\cite{met_eurosys13}).
{\proj}'s current linear auto-scaling policy works well since deploying and invoking new functions is fast;
a more aggressive auto-scaling policy may result in lower SMS capacity utilization with a higher monetary cost. 

\vspace{-6pt}
\subsubsection{Daemon-side Versioning and Persistent Buffer}
\label{subsec:persistent_buffer}

A persistent buffer is a stream buffer that intercepts the chunk streaming on the data path of {\kvput} requests and temporarily buffers the intercepted data on the daemon-local disk. {\proj} uses the daemon-side persistent buffer to accelerate {\kvput} requests without compromising durability and strong consistency. 
Since object chunks written to SMS and COS are read-only, the client daemon uses versioning to support updates. 
{\proj} uses the consistency increasing algorithm proposed in SCFS~\cite{Bessani2014} to achieve strong consistency atop an eventually consistent COS, e.g., AWS S3.
Strong consistency requires a {\kvput} request to return after object chunks have been stored in both SMS and COS. Since {\proj} uses an inexpensive, slow COS layer, the tail latency of {\kvput} requests is not guaranteed if the application must wait for the data to be fully stored in COS. With the persistent buffer, a {\kvput} request can return immediately after being stored in SMS. The client daemon guarantees that the chunks will be successfully stored in COS later by retrying the request using the data stored in the persistent buffer. After the data is stored in COS, its persistent buffer can then be released. For performance, a read-after-write {\kvget} request can be directly served from the persistent buffer if the chunk's buffer has not been released.

\vspace{-6pt}
\subsubsection{Serving {\kvget}s}
\label{subsec:gets}

{\kvget} requests are routed from the daemon to SMS and served by the functions storing chunks.
The client daemon marks chunks of a hot object that has been re-accessed and may trigger asynchronous migration that compacts the chunks to the latest GC-bucket. This compaction is performed by loading chunks from the COS into SMS' latest GC-bucket.
A {\kvget} hits on a chunk in a degraded GC-bucket must trigger the data migration, see Figure~\ref{fig:time_mgmt}(c).
The client daemon performs the compaction in multiple rounds: in each round, the daemon randomly picks a subset (e.g., 50\%) of all marked chunks and migrates the subset to the latest GC-bucket until all chunks are migrated.
The daemon is configured to bound the maximum compaction interval for each compaction operation to prevent compaction bursts from consuming too many SMS resources. There is still a chance that a {\kvget} request may hit on an object with all or part of its chunks residing in a removed function. In this case, the daemon performs synchronous, on-demand migration, which restores cold chunks from COS to the latest GC-bucket.
The daemon updates the mapping table after completed the migration operation. 

\noindent\textbf{Auto-Scaling under {\kvget}s.} 
Elastic demand caching is triggered when a function's request queue is full and the function can not serve more requests. {\kvget}-triggered auto-scaling follows the algorithm in Figure~\ref{fig:placement} and launches more functions---called cache functions---if the {\kvget} throughput surges.  
In {\small\texttt{TestAndPlace()}},
the client daemon makes an on-demand cache request with the following steps:
(1)~check if the function's request queues are full; 
(2)~check if the memory {\small\texttt{HARDCAP}} of the function is reached; 
(3) if these two checks return true, the function caches the requested chunk from COS into function-memory and returns the chunk. 
Demand-cached data under this temporary caching scheme can be evicted to make room for later caching operations or regular object chunks inserted via {\kvput}.
Function-memory space management is discussed in \cref{subsec:buffering}.

\vspace{-6pt}
\subsubsection{Request Queues and Handling Large Objects}
\label{subsec:queue}

{\proj}'s client daemon adopts a two-queue scheme for each function, where each queue is associated with a network connection that connects the client daemon with a function instance.
This scheme separates small requests from large requests in order to avoid the convoy effect in which large requests block small requests.

{\proj} supports large objects by splitting objects into smaller fragments no larger than 200 MB. We implement a data streaming protocol to pipeline object fragments transfer, similar to HTTP/2’s multiplexing~\cite{http2}, to improve transmission efficiency.

\vspace{-6pt}
\subsection{Function-Memory Space Management}
\label{subsec:buffering}

As mentioned earlier, the demand for extra cache functions is temporary and driven by bursty requests. The cache functions will not be fully utilized for storing regular object chunks until all older FGs become full and sealed. Simply discarding cache functions from SMS would impact performance, as a burst of requests may arrive again anytime. To utilize the memory space of the cache functions, {\proj} divides the memory of a function into two partitions, a \emph{storage partition} and a \emph{cache space}: regular object chunks are stored in the storage partition, while the cache space is designated for buffering demand-cached chunks temporarily. There are two categories of demand-cached chunks: hot chunks cached for serving bursty {\kvget}s (\cref{subsec:gets}); and cold chunks from released function-memory (e.g., GC-bucket releasing) temporarily cached  for serving potential out-of-working-set {\kvget}s.
The size of chunks in the cache space is not included in a function's memory consumption, and therefore, chunks can be evicted if
the function's memory is inadequate 
to serve a new {\kvput}. The cache space is also persisted to COS and will be recovered if a cache function is reclaimed (\cref{subsec:parallel_recovery}).

\vspace{-6pt}
\subsection{Fault Tolerance and Data Recovery}
\label{subsec:recovery}

While AWS provides transient function caching that allows a function instance to be used for multiple invocations~\cite{lambda_context}, it does not guarantee an instance will be cached and a cached instance can be reclaimed anytime. Hence, {\proj} needs to be robust against frequent reclamations of function instances.

One possible approach to enabling durability at the SMS level is to replicate each object in the memories of multiple FGs~\cite{infinicache_fast20}. However, this approach would significantly increase the monetary cost (more than $50\%$ of the monetary cost of {\prelim} is for maintaining replica instances) and reduce the effective memory capacity (half of the memory is used for storing replicas when using 2-way replication).
Furthermore, in the highly unreliable environment that is created when users are given no direct control over AWS' internal reclamation policy,
durability is still not guaranteed when replicas are used, since all instances that store replicas may get reclaimed by AWS.

{\proj} takes a different approach to handle frequent reclamation of function instances: all objects are backed up in an inexpensive, persistent COS layer. 
Upon an invoked instance detecting a partial or complete data loss (e.g., an instance has been reclaimed), this instance recovers, from COS, all of the objects previously stored in the reclaimed instance.
Recovering all objects can be slow given instance's bandwidth limit.
Inspired by RAMCloud's fast recovery mechanism~\cite{fast_recovery_sosp11}, {\proj} by default maintains a single copy for each chunk (replication may exist in cache space) in SMS and uses tens of recovery functions, each recovering a portion of all the lost chunks in parallel.
The unique challenge and the differences from RAMCloud are discussed at the end of  \cref{subsec:parallel_recovery}.


\vspace{-6pt}
\subsubsection{Failure Detection}
\label{subsec:failure-detection}

In order to initiate a data recovery, {\proj} needs to establish that a previously invoked instance has been reclaimed since its last invocation. {\proj} does this by using an \textit{insertion log} to keep track of {\kvput} operations for each individual function. The log is used to determine whether an instance's memory state is up-to-date.

\begin{figure}[h]
\begin{center}
\vspace{-10pt}
\includegraphics[width=0.44\textwidth]{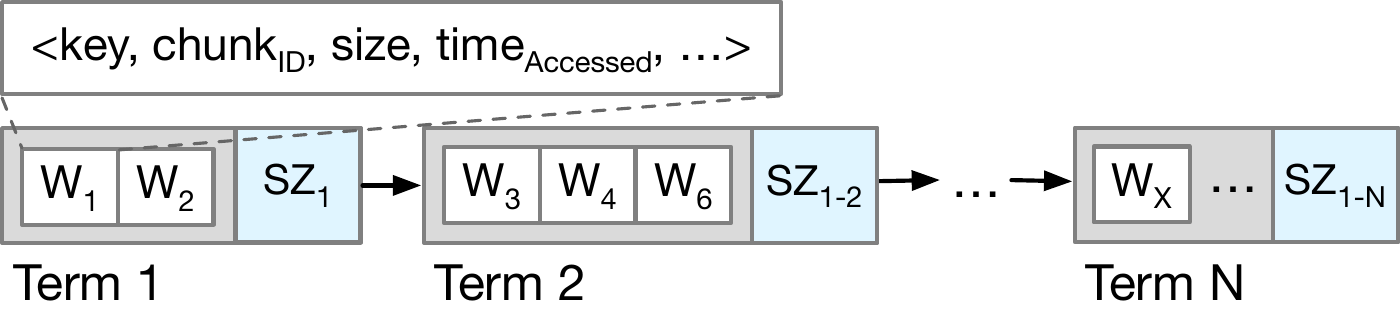}
\vspace{-15pt}
\caption{An example insertion log. $W_1$ denotes the first {\kvput} record stored under {\small\texttt{Term 1}}.
}
\label{fig:lineage}
\vspace{-10pt}
\end{center}
\end{figure}

\noindent\textbf{Insertion Log.}
The insertion log contains one or more insertion nodes, each of which is a COS object that records the {\kvput} operations served by a function instance during a single invocation of the function instance. 
Each {\kvput} results in the creation of an insertion node that is stamped with a monotonically increasing counter value called \emph{term}.
Upon a \kvput, the object chunks are inserted into SMS and then pushed to COS; in the meantime, the function consolidates and seals concurrent PUT records (received at roughly the same time in a time window, whose duration is configurable)
generated during its current invocation into an insertion log node and persists the log node to COS before returning.
Figure~\ref{fig:lineage} shows an example insertion log and the information recorded in each {\small\texttt{PUT}} request.
On returning, the function instance creates a \emph{snapshot} of the chunks it currently stores to speed up recovery downloading.
The snapshot is also persisted in COS. 
The most up-to-date insertion node information is piggybacked onto the {\kvget}/{\kvput} response payload sent to the client daemon.
This information includes the {\small\texttt{term}} of the insertion node, a {\small\texttt{hash}} that is computed from
the insertion node, the {\small\texttt{diff\_rank}}, which is discussed in the next paragraph, the size of the last insertion node stored in COS, and a copy of snapshot information containing all of the aforementioned fields. 
The client daemon maintains up-to-date insertion information for each function it manages.

\noindent\textbf{Triggering a Recovery.}
On a function invocation for a {\kvget}/{\kvput}, the invoked instance $N$ uses the insertion log information passed in the invocation parameter to determine whether data recovery is required. 
AWS Lambda does not guarantee that an invocation of a function constantly reuses the function instance $P$ previously invoked. If $P$ is not reused, the function-memory of $N$ will not contain the most up-to-date object chunks, and these missing chunks will have to be recovered from COS.

An invoked function instance determines whether its objects are up-to-date by performing a consistency check that compares the local {\small\texttt{term}} and {\small\texttt{hash}} values that it has recorded with the corresponding values passed from the client daemon.
If the values are inconsistent, the instance is considered to have \emph{failed}.

When a function instance fails, {\proj} needs to decide on an appropriate recovery strategy.
Specifically, each instance maintains a {\small\texttt{diff\_rank}} that indicates how many object chunks will be recovered since the first {\small\texttt{term}}, which equals the number of all {\kvput}s including deleted chunks. A difference is calculated by subtracting the local {\small\texttt{diff\_rank}} from the {\small\texttt{diff\_rank}} received from the client daemon.
If the difference is significantly larger than $N$, where 
$N$ equals to the number of recovery functions assigned to the function, the failed instance notifies the client daemon that a \emph{parallel recovery} is required (\cref{subsec:parallel_recovery}). Otherwise, the lost data can be recovered locally, and no further action from the client daemon is required.

{\proj} uses the number of chunks rather than the aggregated chunk size as the indicator for parallel recovery because the recovery process can take advantage of the available massive parallelism of recovery functions only if the number of missing chunks is large enough. 
Regardless of whether a parallel recovery is triggered, the failed function instance will perform recovery locally.
The failed instance recovers data by first downloading an operation manifest of the function from COS, which is a combination of a chunk list covered by the last snapshot and operations in the insertion nodes constructed since the latest snapshot, if any.
The instance then replays all operations to find objects to be recovered and downloads objects from COS. Though {\kvget} requests will be blocked until the object has been downloaded, the request latency will not be affected, as discussed in \cref{subsec:parallel_recovery}. To guarantee {\kvput} consistency, the being-recovered storage instance still serves {\kvput} requests, and any successive read-after-write {\kvget}s that
request the chunks inserted by these new {\kvput}s during the
recovery process.

\vspace{-10pt}
\subsubsection{Parallel Recovery}
\label{subsec:parallel_recovery}
\vspace{-6pt}

If parallel recovery is required, the failed function instance notifies the client daemon to start the parallel recovery process. The process involves three phases: recovery group selection, chunk recovery, and service resumption.

\noindent\textbf{Phase 1:  Recovery Group Selection.} 
Each function in the SMS is assigned one of two roles. The role of a \emph{storage} function is to store data objects in function-memory. The role of a \emph{recovery} function is to mitigate the recovery phase's impact by recovering a part of all data missing from the failed storage function, in massive parallel, and delegating {\kvget} requests before the restoration of failed function.
Each storage function is initialized with a group of recovery functions, which are randomly chosen from all the functions in the SMS (\cref{subsec:sms}). The client daemon guarantees that each function may only serve as a recovery function for one storage function at any time. 
The daemon maintains a non-recovering function pool within the active GC-buckets. Upon starting a parallel recovery process, if any previously initialized recovery function is not available (i.e., serving another storage function), a new function will be selected from the pool to serve as the recovery function.

\noindent\textbf{Phase 2: Chunk Recovery.} Each recovery instance is assigned a unique ID $i$ and is responsible for recovering 
a portion of all the chunks to be recovered, which are
the chunks with hashed key $j$ if $j$ modulo the size of the recovery group equals $i$. Recovery instances execute the same recovery routine as described for the failed instance with the exception that they only download objects they are responsible. 

To serve {\kvget} requests for chunks that are being recovered, the client daemon reroutes requests to the corresponding recovery functions.
In \cref{subsec:eval_recovery}, we show that parallel recovery instances will recover a $3,008$ MB function within 1.18 s on average. With erasure coding, the client daemon can tolerate the loss/delay of up to the number of $p$ parity chunks, which greatly reduces the possibility that the instance reclamation impacts the latency of {\kvget} requests.
{\kvput} requests are served by the failed instance, as discussed previously.

\noindent\textbf{Phase 3: Service Resumption.}
Once the failed storage function instance finishes recovering all missing chunks, it notifies the client daemon, which then seamlessly redirects all further {\kvget} requests back to the recovered storage instance. The recovered chunks in the recovery function instances are retained for a certain period before being freed in case a parallel recovery for the same storage function is triggered again in the near future. 

Note RAMCloud~\cite{fast_recovery_sosp11} assumes normal datacenter server failure rates, spreads recovered data across all the recovery nodes, and serves recovered data from recovery nodes permanently. In contrast, {\proj} deals with a more dynamic and unpredictable FaaS environment with higher failure frequencies. To prevent cascading parallel recoveries caused by recovery functions being reclaimed during the recovery of a storage function instance, a recovery function instance only stores recovered data objects temporarily to mitigate the impact of storage instance recovery on request latency.

\vspace{-6pt}
\section{Evaluation}
\label{sec:eval}

In this section, we evaluate {\proj} on AWS Lambda.

\noindent\textbf{Implementation.} We have implemented a production-quality prototype of {\proj} atop {\prelim} by modifying and adding 21,397 lines of Go code~\cite{cloc} using about two person-years: 1,171 LoC for the client library, 10,296 for the client daemon, 6,429 for the Lambda runtime, and 3,501 for utilities shared across components.

\noindent\textbf{Client Setup.} Unless otherwise specified, we deployed the workload replayer and microbenchmarks together with a {\proj} client daemon on a {\small\texttt{c5n.4xlarge}} EC2 VM instance.

\noindent\textbf{Goals.} 
We answer the following questions in the evaluation:
\begin{itemize}[noitemsep,leftmargin=*]
    \item How well does {\proj} elastically adapt to real-world production storage workload changes (\cref{subsec:applications})?
    
    \item Is {\proj} cost-effective and pay-per-access (\cref{subsec:eval_docker})?
    
    \item \addVLDBcomment{How does {\proj} perform under YCSB~\cite{ycsb_socc10} stress testing
    compared to state-of-the-art cloud storage systems (\cref{subsec:eval_ycsb})?}
    
    \item How fast does {\proj} scale out and react to throughput changes \addVLDBcomment{compared to state-of-the-art cloud storage systems (\cref{subsec:eval_elasticity})}? 
    
    \item How fast can {\proj} recover from function failures (\cref{subsec:eval_recovery})?
    \item How much do different design options contribute to {\proj}'s cost-effectiveness and latency improvement (\cref{subsec:eval_factor})?
\end{itemize}

\vspace{-6pt}
\subsection{Applications}
\label{subsec:applications}

\begin{figure}[t]
\begin{center}
\begin{minipage}{\textwidth}

\begin{minipage}[b]{0.23\textwidth}
\includegraphics[width=1\textwidth]{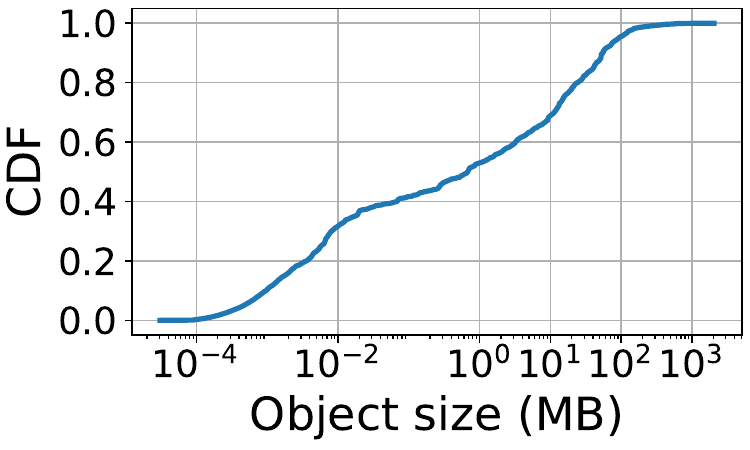}
\vspace{-20pt}
\caption{IBM container registry workload.}
\label{fig:ibm_size_dist}
\end{minipage}
\hspace{.5pt}
\begin{minipage}[b]{0.23\textwidth}
\includegraphics[width=1\textwidth]{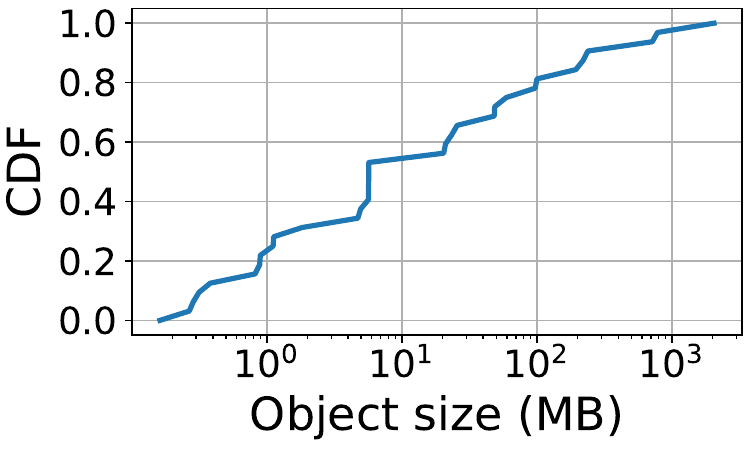}
\vspace{-20pt}
\caption{Azure Functions blob workload.}
\label{fig:azure_size_dist}
\end{minipage}

\end{minipage}
\end{center}
\vspace{-15pt}
\end{figure}


\begin{figure}[t]
\begin{minipage}{\textwidth}

\begin{minipage}[b]{0.245\textwidth}
\includegraphics[width=1\textwidth]{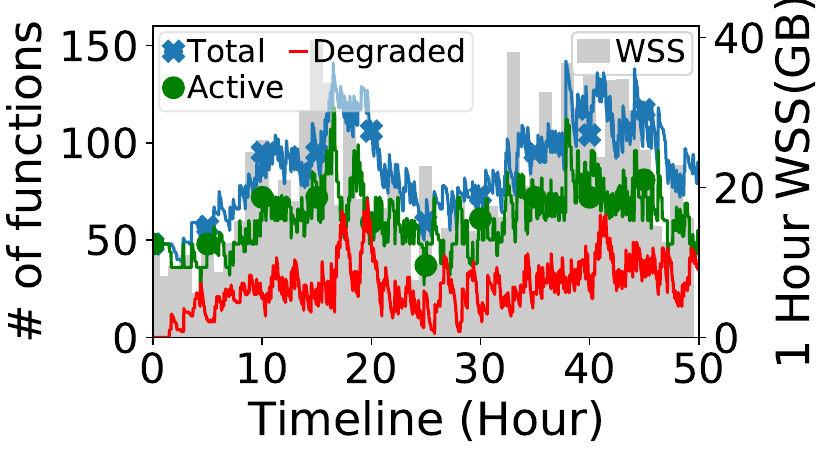}
\vspace{-20pt}
\caption{
Number of functions managed by {\proj}.\\\hspace{\textwidth}
Total=active+degraded. 
}
\label{fig:func_timeline}
\end{minipage}
\hspace{.5pt}
\begin{minipage}[b]{0.22\textwidth}
\includegraphics[width=1\textwidth]{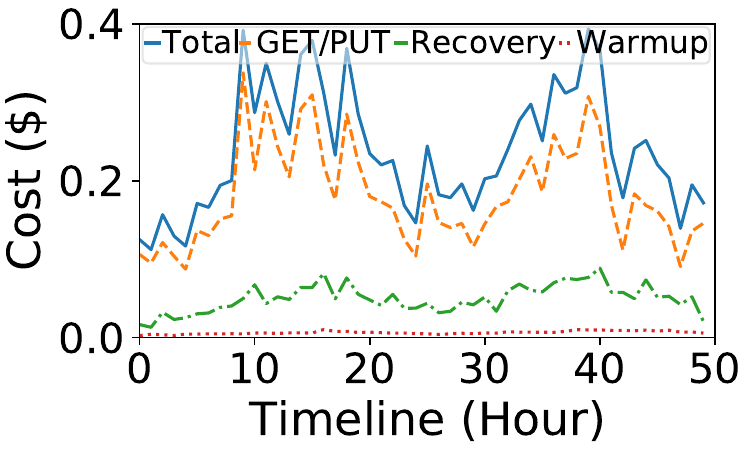}
\vspace{-20pt}
\caption{
Timeline of hourly cost.\\\hspace{\textwidth}
Total=I/O+recovery+warm.
}
\label{fig:cost_timeline}
\end{minipage}

\end{minipage}
\vspace{-10pt}
\end{figure}

\subsubsection{IBM Container Registry Workload}
\label{subsec:eval_docker}

We now evaluate {\proj}'s elasticity, cost, and performance using 
the production IBM container registry  workload\footnote{\cref{sec:moti} describes workload statistics in detail.}. The original workload contains a 75-day request trace collected from 7 geographically distributed datacenters. We use the first 50 hours of the workload from the Dallas datacenter, which features the highest load (with an average throughput of $3,654$ {\kvget} requests per hour). The distribution of object sizes is shown in Figure~\ref{fig:ibm_size_dist}. Among all objects, $31\%$ are larger than 10 MB.
We compare our results with {\prelim}~\cite{infinicache_fast20} and AWS ElastiCache for Redis~\cite{aws_ec}. 
Following workload configurations are applied to both {\prelim} and {\proj}:

\begin{itemize}[noitemsep,leftmargin=*]
    \item Parse all the {\kvget} requests reading a container image layer. 
    \item Include all objects (both small and large objects $>$ 10 MB).
    \item Use a Reed-Solomon EC configuration of $(10+2)$. 
    \item Use Lambda function instances with $1,536$ MB memory.
\end{itemize}

For {\prelim}, we use a fixed cluster of 400 Lambda functions and a warmup interval of 1 minute.
{\proj} applies a warmup interval of 1 minute for Lambda functions in active GC-buckets and a longer warmup interval of 5 minutes for degraded GC-buckets. The GC interval is set to 10 minutes. The number of active ($M$) and degraded ($N$) GC-buckets are set to 6 and 12, respectively, based on the average reuse interval and the average one-hour WSS.

\begin{figure*}[t]
\begin{minipage}{\textwidth}

\hspace{2pt}
\begin{minipage}[b]{0.32\textwidth}
\begin{center}
\includegraphics[width=1\textwidth]{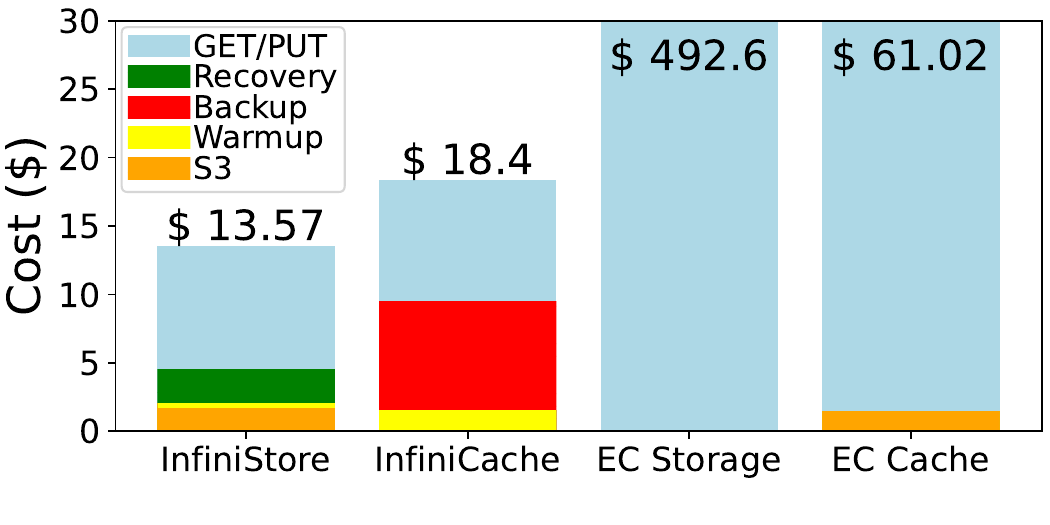}
\end{center}
\vspace{-10pt}
\caption{
\$ cost.
{\small\texttt{EC}}: ElastiCache.
}
\vspace{-6pt}
\label{fig:cost_comparison}
\end{minipage}
\hspace{1pt}
\begin{minipage}[b]{0.32\textwidth}
\begin{center}
\includegraphics[width=1\textwidth]{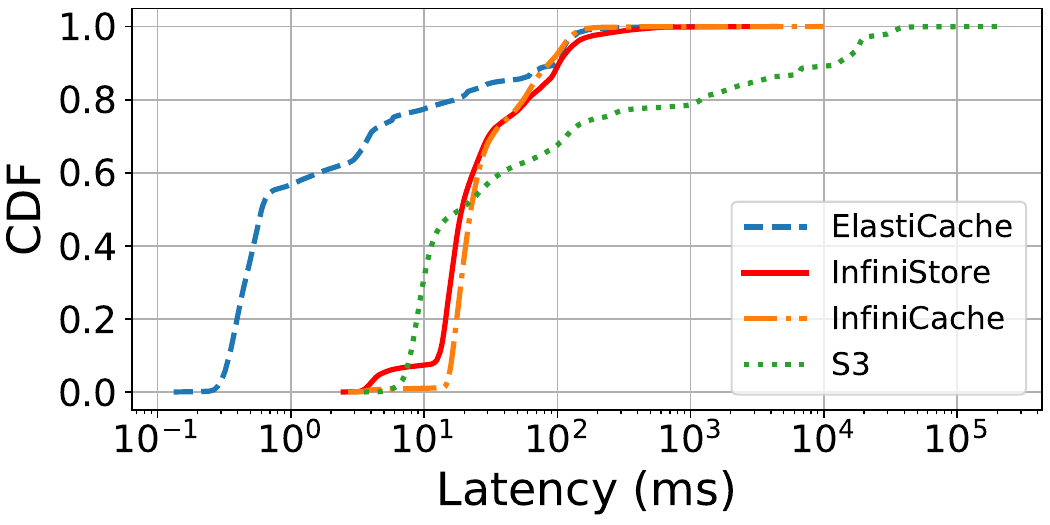}
\end{center}
\vspace{-12pt}
\caption{
Latency comparison (all obj).
}
\vspace{-6pt}
\label{fig:latency}
\end{minipage}
\hspace{-4pt}
\begin{minipage}[b]{0.32\textwidth}
\begin{center}
\includegraphics[width=1\textwidth]{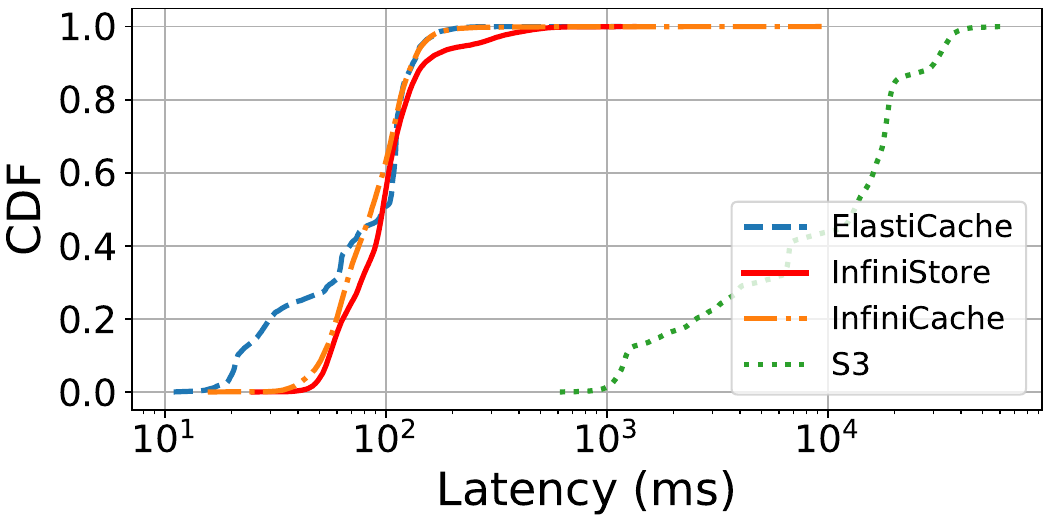}
\end{center}
\vspace{-12pt}
\caption{
Latency comparison (>10 MB).
}
\vspace{-6pt}
\label{fig:latency_10MB}
\end{minipage}

\end{minipage}
\vspace{-12pt}
\end{figure*}

\begin{table}[t]
\begin{center}
\begin{tabular}{lrrrrr}
{\bf Workload} & {\bf IS SMS} & {\bf IC} & {\bf EC S} & {\bf EC C} \\
\Xhline{2\arrayrulewidth}
{\bf IBM container registry} & $95.8\%$ & $95.4\%$ & $99.96\%$ & $93.3\%$ \\
\end{tabular}
\caption{Workload's memory-level read hit ratio achieved by {\proj} ({\small\texttt{IS}}), {\prelim} ({\small\texttt{IC}}), ElastiCache as a storage ({\small\texttt{EC S}}), and ElastiCache as a cache ({\small\texttt{EC C}}). 
{\proj}'s read hit ratio is defined as the ratio of the number of object chunks read directly from the SMS layer against the total object chunk requests.}
\label{tbl:hit_ratio}
\vspace{-25pt}
\end{center}
\end{table}

\begin{figure*}[ht]
\begin{center}
\subfigure[1 MB objects.] {
\includegraphics[width=.32\textwidth]{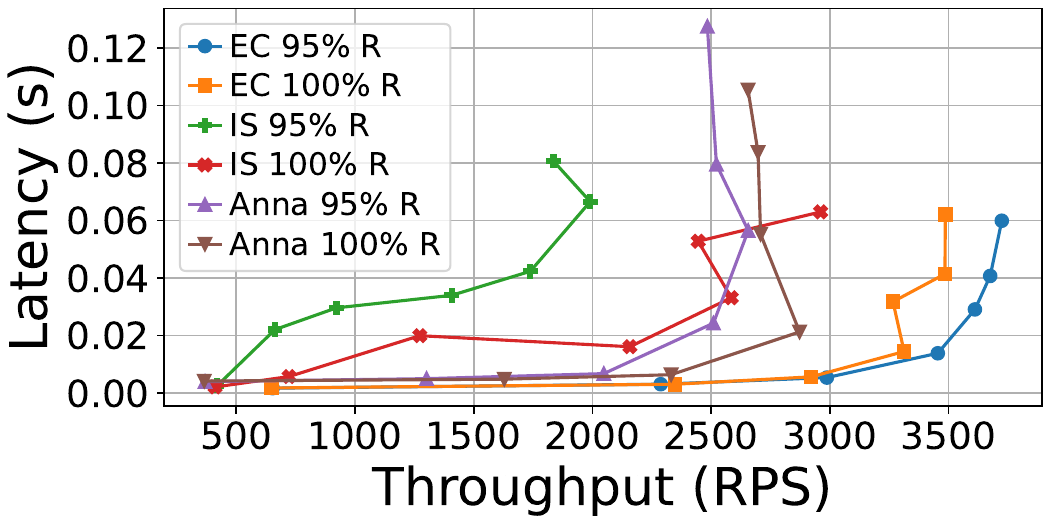}
\label{fig:ycsb_1M}
}
\hspace{-7pt}
\subfigure[10 MB objects.] {
\includegraphics[width=.32\textwidth]{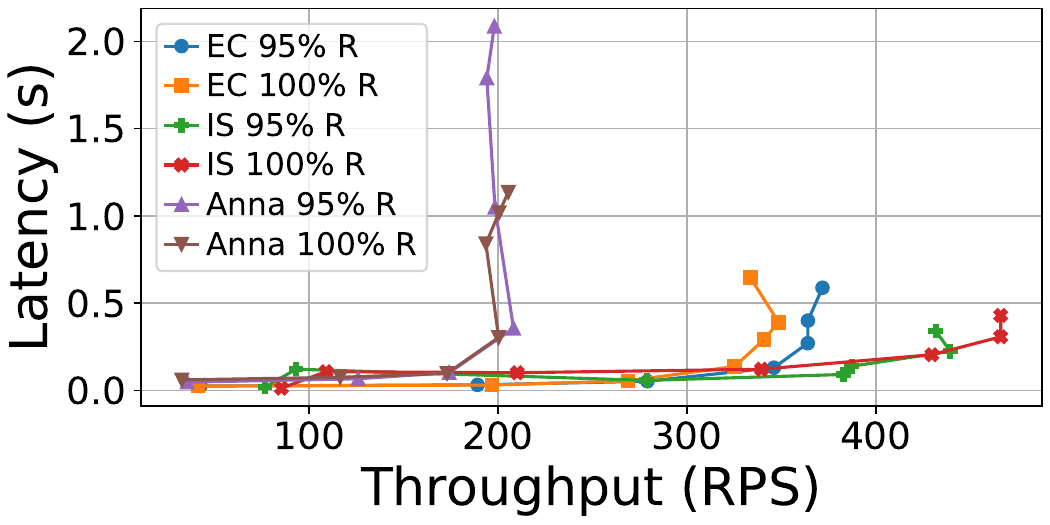}
\label{fig:ycsb_10M}
}
\hspace{-7pt}
\subfigure[100 MB objects.] {
\includegraphics[width=.32\textwidth]{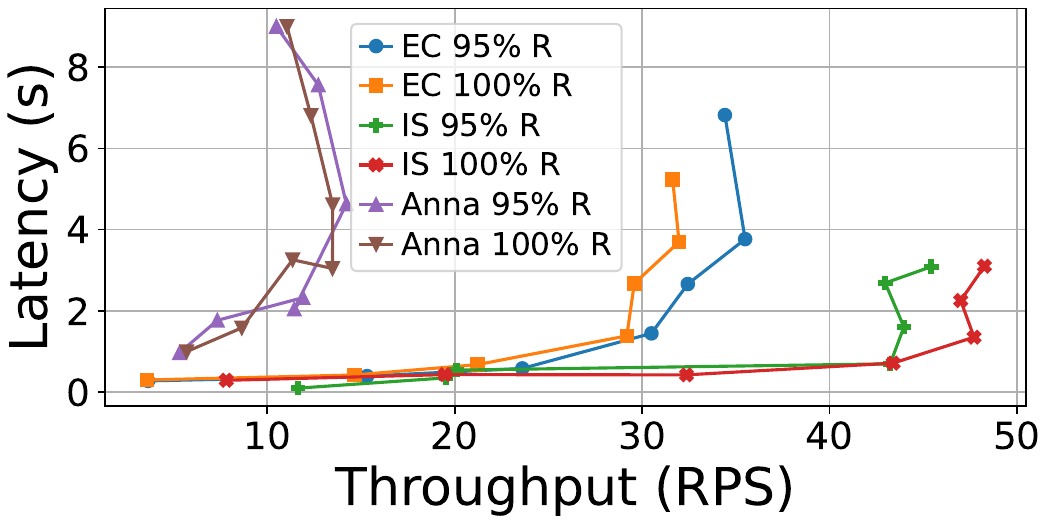}
\label{fig:ycsb_100M}
}
\vspace{-15pt}
\caption{
\addVLDBcomment{Throughput vs. the p90 read latency obtained under YCSB with various object sizes. {\small\texttt{IS}}: {\proj}.}
}
\label{fig:ycsb}
\end{center}
\vspace{-10pt}
\end{figure*}


\noindent\textbf{Elasticity.}
We start by evaluating {\proj}'s ability to adapt to working set size changes (Figure~\ref{fig:workload_wss}).
Figure~\ref{fig:func_timeline} shows the number of functions managed by {\proj} over the 50-hour workload. Each curve corresponds to the number of functions managed in {\proj}'s active GC-buckets (green) and degraded GC-buckets (red) at each 1-minute time interval, regardless of whether the function instances are invoked or not. Note that the number of total functions matches 1 hour WSS changes, \diffVLDBcomment{which are capped at 40 GB out of an aggregate 888 GB of workload's 50 hours WSS.}{(Aggregate 50 hours WSS of the workload is 888 GB.)}
\addVLDBcomment{The measured SMS-level hit ratio is $95.8\%$ (see Table~\ref{tbl:hit_ratio}), which indicates that $95.8\%$ of requests are directly served from SMS without loading data from COS. The high SMS-level hit ratio also indicates that {\proj} can automatically capture application's working set.}

\noindent\textbf{Cost-effectiveness.}
Next, we evaluate the overall monetary cost of {\proj} in comparison with {\prelim} and AWS ElastiCache for serving the container registry workload (Figure~\ref{fig:cost_comparison}). For ElastiCache, we provisioned two Redis clusters: a cluster of 12 {\small\texttt{cache.r6g.2xlarge}} instances with 633.64~GB aggregate memory to provide full in-memory storage and a cluster of 8 {\small\texttt{cache.m6g.large}} instances with 51.04~GB aggregate memory to provide in-memory caching. For fairness, the cache cluster uses S3 as the backing store and has reduced hit ratio (Table~\ref{tbl:hit_ratio}).
Both ElastiCache clusters have up to 10~Gbps network bandwidth per instance.

{\prelim} costs a total of $\$18.4$, of which $43.48\%$ is spent on the backup scheme. ElastiCache costs $\$492.6$ (as a storage) and  $\$61.02$ (as a cache with S3 as the backing store), $36.30\times$ and $4.50\times$ more expensive than {\proj} as it is statically provisioned. {\proj} has the lowest cost of $\$13.57$, of which $\$2.49$ is used for parallel recoveries, $\$0.31$ for warmup, and $\$1.68$ for COS (S3) storing 1053~GB of data (including overhead of using erasure coding) in 50 hours.
Note that {\proj}'s warmup cost is considerably lower than {\prelim}, which maintains a fixed-sized function pool with replicas, whereas {\proj} uses a sliding-window scheme to dynamically adjust the number of functions in SMS. 

\noindent\textbf{Pay-per-Access.}
To find out whether \proj delivers pay-per-access, we breakdown the cost per hour over the entire 50 hours (Figure~\ref{fig:cost_timeline}). 
We observe a clear trend in that the total cost is proportional to the cost of serving {\kvget}/{\kvput} requests. {\proj}'s parallel recovery scheme incurs a small portion ($18.34\%$) of the overall cost, which is the recovery cost for handling $1,083$ function instance failures during the workload.

{\proj}'s cost overhead over an ideally pay-per-access scheme is 
$26.00\%$. This result is calculated using the ratio of the aggregate cost of recovery + warmup (i.e., the extra activities required to maintain data durability) to the aggregate cost of serving {\kvget}/{\kvput} requests + S3 cost (i.e., access and storage cost). This overhead is significantly lower than {\prelim}, which adds $106.51\%$ for backup and warmup.

\noindent\textbf{Latency Performance.}
Figure~\ref{fig:latency} compares {\proj}'s latency performance against {\prelim}, ElastiCache as a storage, and S3. The results of using ElastiCache as a cache are similar to use as a storage, which are not shown in plots.
{\proj} achieves lower latency than {\prelim} for more than $40\%$ of the requests. This is because {\proj}'s two-queue request scheme effectively mitigates the convoy effect of large requests against small requests.
{\proj} is significantly faster than S3 for more than $70\%$ of the requests. 
For objects larger than 10 MB, {\proj} is two orders of magnitude faster than S3.
For objects smaller than 1 MB, {\proj} suffers from the overhead of Lambda function invocation and thus is lower than S3.
{\proj} exhibits comparable performance to ElastiCache for around $50\%$ of requests reading objects over 10 MB (Figure~\ref{fig:latency_10MB}). 
For the rest of the requests accessing smaller objects, {\proj} does not see a benefit compared to ElastiCache, again because of the high invocation overhead. Overall, {\proj} offers \emph{a novel performance-\$cost tradeoff} in the space of general-purpose cloud storage services. 

\vspace{-6pt}
\subsubsection{Azure Functions Blob I/O Workload}
\label{subsec:azure_wl}

\begin{figure}[t]
\begin{center}
\vspace{-6pt}
\includegraphics[width=0.42\textwidth]{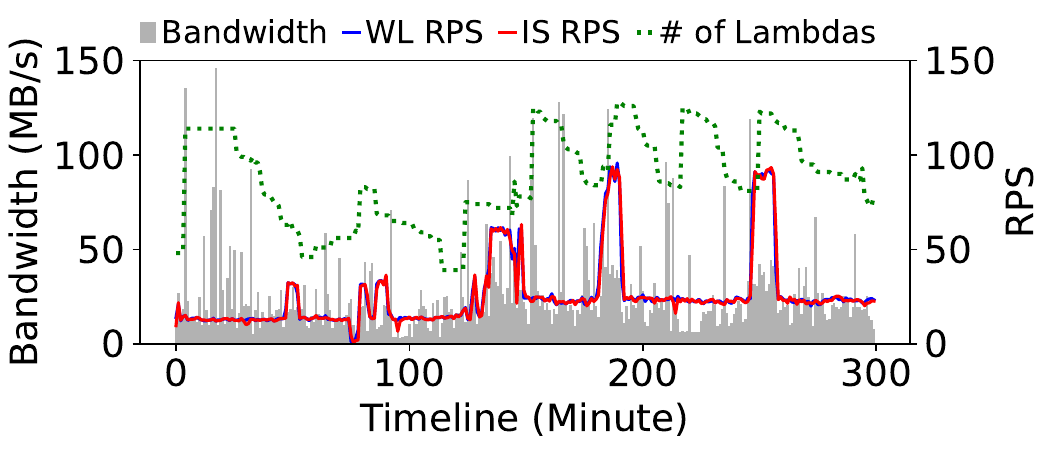}
\vspace{-10pt}
\caption{Timeline of the bandwidth and RPS (request per second) changes of the Azure Functions blob access workload. {\proj}(IS)’s trace replayer faithfully replays  the workload in real-time ({\small\texttt{IS RPS}}) with the same I/O concurrency as recorded in original traces ({\small\texttt{WL RPS}}).
} 
\label{fig:azure_throughput}
\vspace{-22pt}
\end{center}
\end{figure}

The full Azure Functions trace contains 14~days of blob accesses in 855 serverless function applications and is highly-bursty (Figure~\ref{fig:workload_reuse_cov}). We selected a 5-hour trace (hour 179--184), which contains the most bursty I/O pattern (with a mean $CoV > 1$) blob access information issued by a total of 37 function applications. We increased the blob size by $10,000\times$ to highlight how {\proj} elastically scales in response to increasing bandwidth requirements. Figure~\ref{fig:azure_size_dist} shows that $45\%$ of the objects are larger than 10~MB. Since the Azure workload has much shorter object reuse intervals (see Figure~\ref{fig:workload_reuse_cdf}) than that of the container registry workload, we set the GC interval as 1~minute, and the number of active ($M$) and degraded ($N$) GC-buckets as 8 and 15, respectively.
As shown in Figure~\ref{fig:azure_throughput}, {\proj} dynamically adjusts the number of Lambda functions based on the bandwidth requirement and the RPS (requests per second) changes. {\proj} can effectively scale to the bandwidth requirement, thanks to {\proj}'s optimization for handling large objects (\cref{subsec:queue}), which splits large objects into smaller pieces. This way, a large request gets converted into multiple smaller requests that can be quickly absorbed in parallel by scaled out Lambda function instances.

\vspace{-6pt}
\subsection{YCSB Microbenchmarking}
\label{subsec:eval_ycsb}

Next, we stress-test the throughput-latency performance tradeoff using the commonly-used YCSB~\cite{ycsb_socc10} benchmark. We compare {\proj} against two state-of-the-art cloud storage systems: 
(1)~AWS ElastiCache for Redis, 
and (2)~Anna~\cite{anna}, an auto-scaled, multi-tier cloud storage system used by Cloudburst~\cite{cloudburst} to support stateful serverless workloads. 
For both ElastiCache and Anna, we deploy a small serverful cluster consisting of three 8-vCPU, 50-GB-memory EC2 VMs with up to 30~Gbps aggregated, cluster-wide network bandwidth. The ElastiCache deployment and the Anna deployment use slightly different VM instance types due to limited options. Specifically, ElastiCache uses the {\small\texttt{cache.r6g.2xlarge}} cache node type, while Anna uses the {\small\texttt{r4.2xlarge}} EC2 instance type. The Anna deployment uses one additional {\small\texttt{r4.2xlarge}} EC2 instance as the routing server and one {\small\texttt{m4.xlarge}} EC2 instance for management and monitoring services. Both of the two deployments are configured with a replication factor of 1. YCSB runs on a {\small\texttt{c5n.9xlarge}} client VM with a fixed 50~Gbps network bandwidth to ensure that the client's network does not become a performance bottleneck. We configure Anna to use 8 worker threads for each of the three nodes in the storage cluster. {\proj} uses a GC interval of 1 minute. 
For Anna, since there is no Anna-based YCSB binding available, we modified Anna's own benchmarking utility to make it generate YCSB-styled I/O tests. We use two read/update ratios: $95:5$ and $100:0$, and the Zipfian key popularity distribution with a Zipfian coefficient of $0.99$. We run each YCSB test for 30~seconds for the following three object sizes: 1~MB, 10~MB, and 100~MB. For each test, the concurrency (i.e., number of YCSB threads) is configured to start from 1 and then gradually increase to 5, 10, 25, 50, 75, and finally 100. Figure~\ref{fig:ycsb} reports the performance results.

ElastiCache is highly optimized for AWS-hosted I/O workloads in that the YCSB benchmark can easily saturate the aggregated 30 Gbps cluster bandwidth for all three object sizes. ElastiCache also achieves the lowest p90 ($90^{th}$-percentile) latency for the 1-MB-object tests. Anna cannot fully utilize the cluster bandwidth. 
As the object size increases, Anna tends to have a reduced network bandwidth utilization (see Figure~\ref{fig:ycsb_10M}-\ref{fig:ycsb_100M}). To figure out the reason, we investigate Anna's implementation and observe that Anna incurs extra serialization/deserialization overhead when passing large messages through Anna's ZeroMQ service~\cite{zeromq}. Such overhead increases significantly when object sizes increase. {\proj}'s elastic, scaled-out serverless design offers ample network bandwidth resources for applications. That is why (1)~{\proj}'s throughput is only capped by the bandwidth limit of the client VM for both the 10 MB and 100 MB objects,
and (2)~{\proj} achieves the lowest p90 latency compared to both ElastiCache and Anna. For the $100\%$-read, 1-MB-object workloads, {\proj} achieves on-par throughput as Anna, with the highest throughput between 2,500-3,000 RPS and a slightly higher p90 latency (see Figure~\ref{fig:ycsb_1M}). This is because {\proj} fetches data from serverless functions but not a stable serverful cluster. The YCSB results show that {\proj} is well suited for large-object-intensive workloads with object size larger than 10 MB.

\vspace{-6pt}
\subsection{Elasticity Microbenchmarking}
\label{subsec:eval_elasticity}

\begin{figure}[t]
\begin{center}
\begin{minipage}{\textwidth}

\begin{minipage}[b]{0.23\textwidth}
\includegraphics[width=1\textwidth]{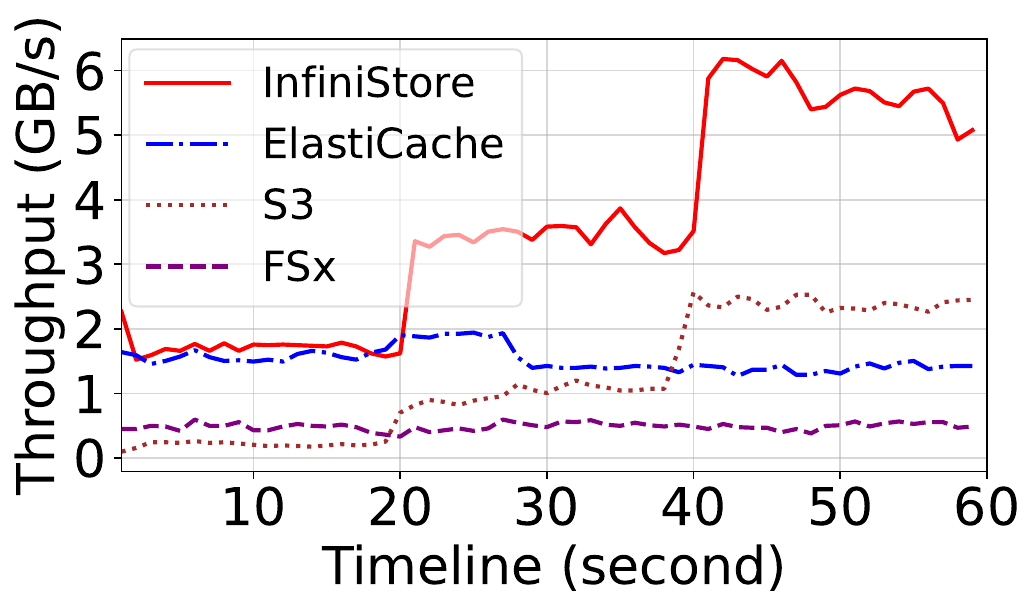}
\vspace{-15pt}
\caption{
Throughput.
}
\label{fig:elastic_thpt}
\end{minipage}
\begin{minipage}[b]{0.23\textwidth}
\includegraphics[width=1\textwidth]{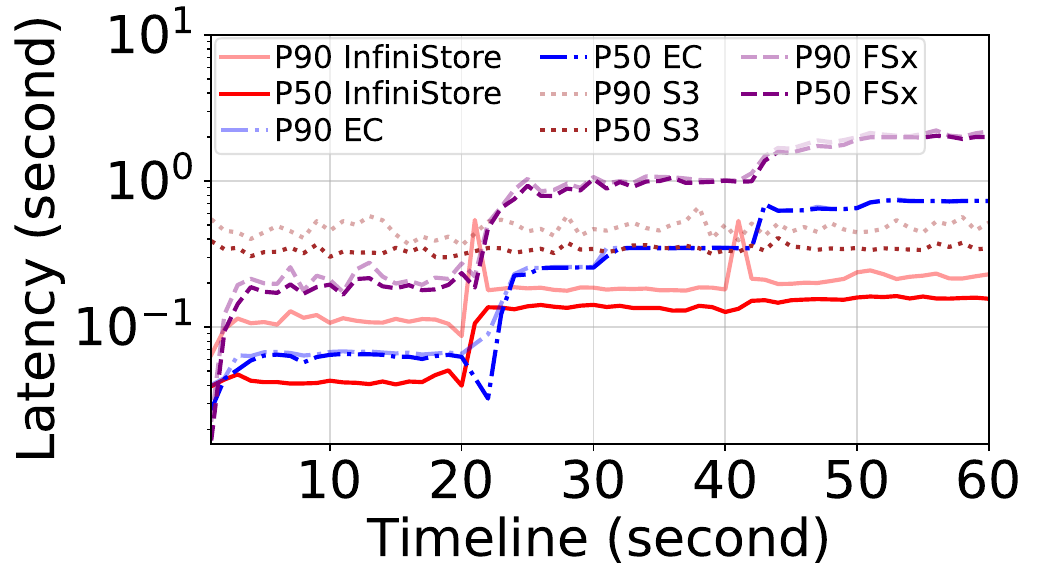}
\vspace{-15pt}
\caption{
Latency (log).
}
\label{fig:elastic_latency}
\end{minipage}

\end{minipage}
\end{center}
\vspace{-15pt}
\end{figure}

In this section, we setup a multi-VM deployment using 10 {\small\texttt{c5n.4xlarge}} EC2 VMs to simulate a realistic use case in which a tenant has multiple microservices that concurrently issue {\kvget} requests for 10 MB objects. 
To evaluate {\proj}'s ability to scale out on demand,
we vary the number of I/O threads on each client VM from 1 to 5 to 10. The load increases every 20 seconds. The WSS of the workload is 211 GB. We compare {\proj} with three cloud storage services, an ElastiCache deployment of one 314.32~GB {\small\texttt{cache.m5.24xlarge}} instance (the smallest ElastiCache instance that provides 25 Gbps network bandwidth), a baseline FSx deployment of 1.2 TB, and S3.
%

Figure~\ref{fig:elastic_thpt} shows that {\proj}'s throughput scales instantly as more clients are added. We can also see that S3 scales but with a much lower throughput. ElastiCache and FSx, on the other hand, hit a network bandwidth bottleneck with a capped throughput of $1.52$ GB/s and $489$ MB/s (FSx provides burst throughput), respectively. 
Both ElastiCache and FSx see a linear increase in latency (Figure~\ref{fig:elastic_latency}), since they both need to be manually re-configured to scale out. 
\addVLDBcomment{At p90, we see a latency spike when the load increases. This is because more functions need to be invoked on demand to sustain the burst. As shown, {\proj} quickly scales out to sustain the latency spike.}
{\proj} achieves $46.72\%$ and $81.84\%$ lower latency at the
$90^{th}$ percentile compared to ElastiCache and FSx, respectively.

%

\vspace{-6pt}
\subsection{Parallel Recovery}
\label{subsec:eval_recovery}

Next, we evaluate the performance and effectiveness of {\proj}'s parallel recovery scheme. 

\begin{figure}[t]
\begin{minipage}{\textwidth}

\begin{minipage}[b]{0.23\textwidth}
\begin{center}
\includegraphics[width=1\textwidth]{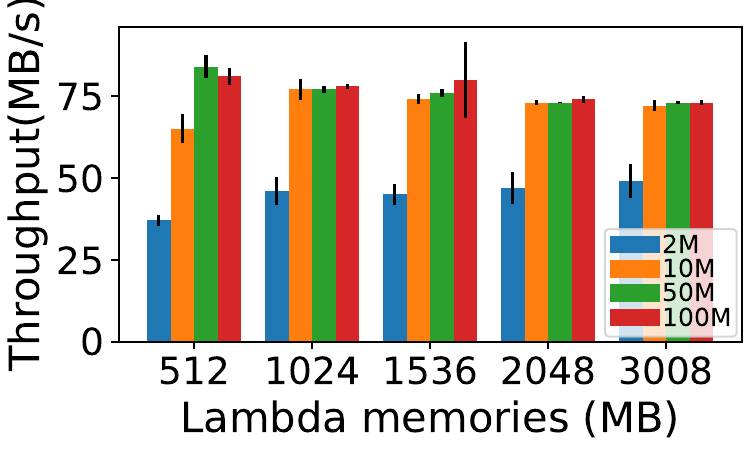}
\end{center}
\vspace{-12pt}
\caption{
$\lambda$ throughput. 
}
\label{fig:instance_thpt}
\end{minipage}
\hspace{4pt}
\begin{minipage}[b]{0.23\textwidth}
\begin{center}
\includegraphics[width=1\textwidth]{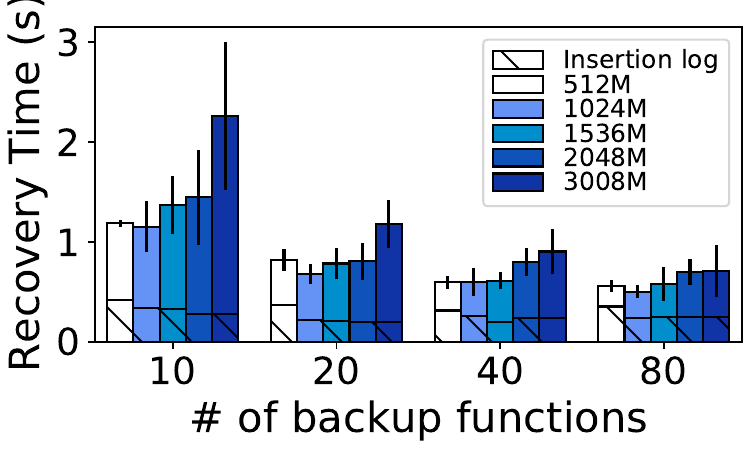}
\end{center}
\vspace{-12pt}
\caption{
Recovery time.
}
\label{fig:recovery_time}
\end{minipage}

\end{minipage}
\vspace{-10pt}
\end{figure}

\noindent\textbf{Lambda Throughput.}
To better understand the limitations of the parallel recovery performance, we examine the network throughput of the Lambda function instances. Each Lambda instance runs 10 threads that concurrently fetch data from S3 and we vary the Lambda function-memory size from 512 MB to 3008 MB while measuring the throughput of downloading objects with sizes ranging from 2 MB to 100 MB. Figure~\ref{fig:instance_thpt} reports the results.  We observe that:
(1)~downloading objects of 2 MB cannot saturate a Lambda's network bandwidth,
and (2)~for all memory configurations, the sustained network throughput can reach up to around 75 MB/s, suggesting that a single recovery function instance is capable of recovering 100 MB of data in 1 or 2 seconds.

\noindent\textbf{Performance of Parallel Recovery.}
To analyze \proj{}'s parallel recovery performance, we measure the time it takes to recover the data objects stored in a Lambda function instance with various memory configurations from 512 MB to $3,008$ MB. As shown in Figure~\ref{fig:recovery_time}, recovering a $3,008$ MB Lambda function instance takes, on average, 1.18 s when using 20 parallel recovery functions. The recovery time is reduced by only $39.9\%$ if the number of recovery functions scales from 20 to 80. Insertion log downloading takes, on average, 271.4 ms of the average recovery time for each recovery function.
To balance the gain and overhead, we choose to use 20 recovery functions in other experiments.

\begin{figure}[t]
\begin{minipage}{\textwidth}

\begin{minipage}[b]{0.234\textwidth}
\begin{center}
\includegraphics[width=1\textwidth]{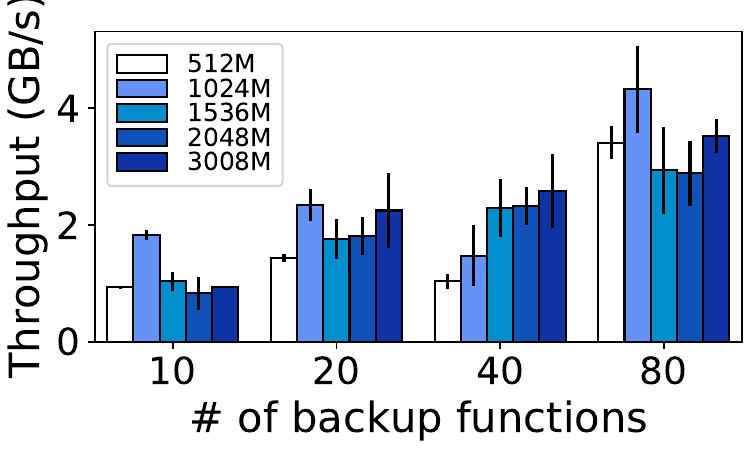}
\end{center}
\vspace{-12pt}
\caption{
Recovery thpt.
}
\label{fig:recovery_thpt}
\end{minipage}
\hspace{4pt}
\begin{minipage}[b]{0.232\textwidth}
\begin{center}
\includegraphics[width=1\textwidth]{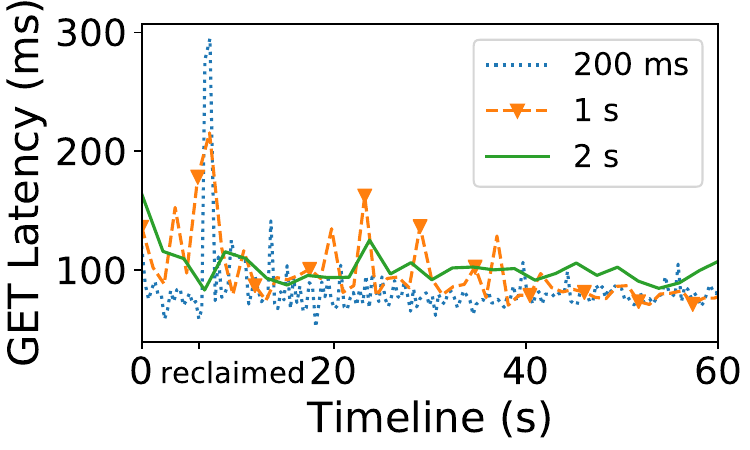}
\end{center}
\vspace{-12pt}
\caption{
Impact on {\kvget}s. 
}
\label{fig:recovery_impact}
\end{minipage}

\end{minipage}
\vspace{-20pt}
\end{figure}

Accordingly, Figure~\ref{fig:recovery_thpt} shows that the aggregate recovery throughput achieved to recover a $3,008$~MB function increases from $2.25$ GB/s to $3.51$ GB/s 
when scaling the recovery group from 20 to 80 functions. Similar trends can be observed for other Lambda configurations and recovery group sizes. 

\noindent\textbf{Impact of Recovery on Latency.}
We next evaluate the impact of parallel recovery on {\kvget} latency.
In this test, we first load $2,400$ unique 10~MB objects into a {\proj} deployment of one FG (Function Group) of 12 $3,008$~MB Lambda instances with an EC code of $(10+2)$. We then run a client that issues random {\kvget} requests in a fixed interval.
We choose three different intervals: 200 ms, 1 s, and 2 s. For each test run,
we kill one Lambda function
5 s after the start of the {\kvget} workload
to simulate a 
reclamation event performed by the provider. On detecting the reclamation of the Lambda instance, {\proj} is configured to invoke a group of 20 independent recovery instances to perform the parallel recovery.

As shown in Figure~\ref{fig:recovery_impact}, the client does not experience any service interruptions during the run with a 2 s request interval. Because {\proj} finishes parallel recovery within 2 s, and the requests for the lost object chunks were seamlessly served by the 20 recovery instances while waiting for the storage instance to fully recover from S3. In the tests with a request interval of 200 ms or an interval of 1 s, we observe a latency increase of about 200 ms. This is because, although 
the build-in erasure coding feature can tolerate losing 2 out of $(10+2)$ chunks, the process of decoding and reconstructing the object has an impact on latency.
However, after parallel recovery, the {\kvget} latency decreases, demonstrating the effectiveness of {\proj}'s parallel recovery scheme.
\vspace{-6pt}
\subsection{Factor Analysis}
\label{subsec:eval_factor}

\begin{figure}[t]
\begin{minipage}{\textwidth}

\begin{minipage}[b]{0.23\textwidth}
\begin{center}
\includegraphics[width=1\textwidth]{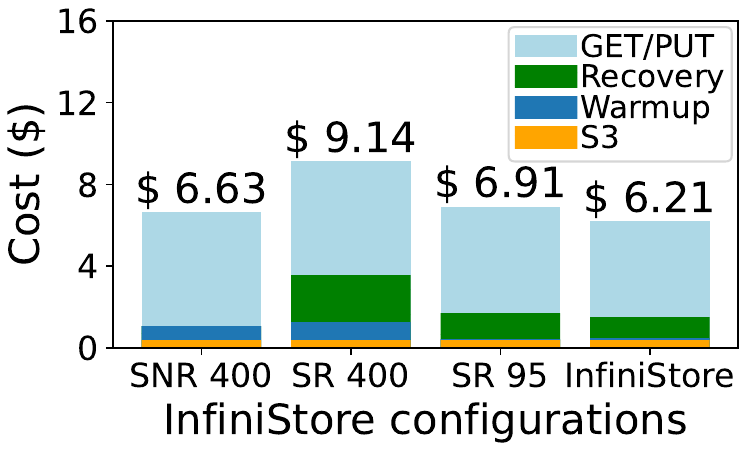}
\end{center}
\vspace{-12pt}
\caption{
Cost of different configurations.
}
\label{fig:factor_cost}
\end{minipage}
\hspace{4pt}
\begin{minipage}[b]{0.23\textwidth}
\begin{center}
\includegraphics[width=1\textwidth]{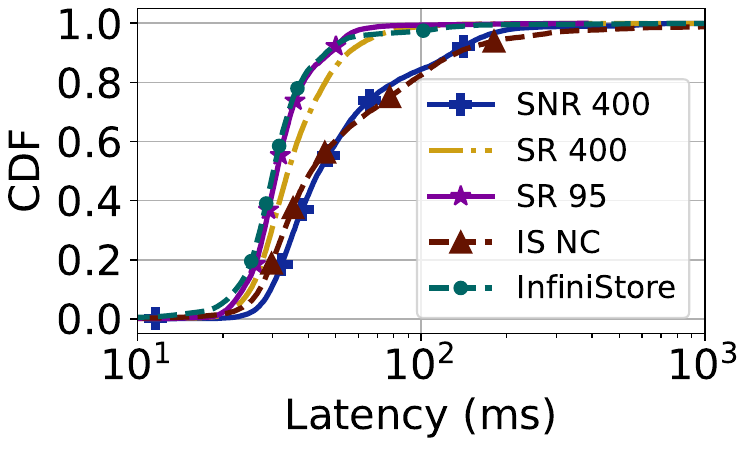}
\end{center}
\vspace{-12pt}
\caption{
Latency of different configurations.
}
\label{fig:factor_latency}
\end{minipage}

\end{minipage}
\vspace{-15pt}
\end{figure}

Finally, we measure how different design options affect {\proj}'s cost effectiveness and latency. We use the following {\proj} configurations: 
(1)~{\small\texttt{SNR 400}}: a static cluster of 400 Lambda functions with no parallel recovery;
(2)~{\small\texttt{SR 400}}: a static cluster of 400 Lambda functions with parallel recovery to emulate a static {\prelim} setup;
(3)~{\small\texttt{SR 95}}: a static cluster of 95 Lambda functions with parallel recovery; the rationale of using 95 functions is that we observe that {\proj} uses an average of 95 functions per minute during the 24-hour workload, though with dynamic adaptations throughout; thus, for comparison, we also test the baseline {\small\texttt{SR 95}} to highlight the benefits brought by {\proj}'s sliding-window management;
(4)~{\small\texttt{IS NC}}: {\proj} with no temporary cache functions for serving bursty request;
(5)~{\proj} with all features enabled. 
We drive the tests using a 24-hour IBM container workload with all workload configurations the same as used in \cref{subsec:eval_docker}. Figure~\ref{fig:factor_cost} and \ref{fig:factor_latency} report the results. {\small\texttt{IS NC}}'s cost is omitted in Figure~\ref{fig:factor_cost} as it is almost the same as that of {\proj}. Figure~\ref{fig:factor_latency} omits the latency results of objects larger than 10~MB as they show negligible differences across all configurations.

\noindent\textbf{Impact of Parallel Recovery.}
As shown in Figure~\ref{fig:factor_latency}, {\small\texttt{SNR 400}} observes increased latency as data objects lost due to function reclamation must be fetched from S3. Interestingly, {\small\texttt{SNR 400}} costs even more than {\proj} despite the smaller cost overhead of warmup  (Figure~\ref{fig:factor_cost}): {\small\texttt{SNR 400}} spends more money serving {\kvget}/{\kvput} requests as significantly more SMS misses lead to prolonged Lambda execution time for fetching missed objects from S3.

\noindent\textbf{Impact of Sliding-window Management and Cache Functions.}
{\proj} dynamically adjusts SMS capacity based on the workload's WSS, while {\small\texttt{SR 95}} uses a fixed function instance pool whose capacity is on average slightly larger than the actual WSS: {\small\texttt{SR 95}} costs $11.3\%$ more than {\proj} (Figure~\ref{fig:factor_cost}). 
Recall that temporary cache functions are launched to serve request bursts (\cref{subsec:puts} and \cref{subsec:gets}). {\small\texttt{IS NC}} has an object chunks miss rate of $18.3\%$ at the SMS layer with cache functions disabled, compared to a much lower miss rate of $1.73\%$ under {\proj}.
As a result, we see from Figure~\ref{fig:factor_latency} that {\small\texttt{IS NC}} suffers a big latency penalty.
\vspace{-6pt}
\section{Related Work}
\label{sec:related}

\noindent\textbf{Serverless Data Services.} Fully-managed cloud storage services~\cite{aws_s3, gcs, azure_blob, aws_aurora_sigmod18, aws_dynamodb} 
transparently manage storage resources for tenants. 
Researchers have also explored offloading data services to serverless functions.
Starling~\cite{starling_sigmod20} supports database query processing using serverless functions. 
Zion~\cite{zion_middleware17} extends object storage by offloading data processing to stateless serverless functions.
Unlike {\proj}, these systems do not directly store data in functions, which presents new challenges that {\proj} addresses. 

\noindent\textbf{Caches/Storage for Serverless Applications.}
OFC~\cite{Mvondo2021} is an opportunistic memory cache that leverages overbooked host memory of serverless functions to accelerate function execution. Faa\$T~\cite{faast_socc21} is a distributed cache co-located with FaaS applications. Faa\$T scales as the application scales.
Pocket~\cite{pocket_osdi18, ephemeral_atc18} is an elastic ephemeral storage for serverless analytics~\cite{pywren_socc17, locus_nsdi19, wukong_pdsw19, wukong_socc20}. SONIC~\cite{sonic_atc21} optimizes data exchange among chained functions.
AFT~\cite{aft_eurosys20} is a fault-tolerant shim providing atomic transaction guarantees to serverless applications that store data in cloud storage. 
Shedder~\cite{Zhang2019} moves function computation directly to cloud storage.
\addVLDBcomment{Anna's lattice support can achieve coordination-free causal consistency~\cite{anna}. Cloudburst~\cite{cloudburst} adds host-local cache atop Anna to support stateful function pipelines.} \textsc{FaaSNet}~\cite{faasnet_atc21} accelerates container provisioning for serverless functions using a P2P tree.
In contrast, {\proj}'s SMS layer leverages serverless functions to achieve rapid \addcomment{and find-grained} storage-level scaling 
and is designed to serve the I/Os of general cloud applications.

\noindent\textbf{Cost-effective Cloud Storage.}
Prior works exploit the performance and cost heterogeneity of a combination of cloud storage services (e.g., VM-local block drives, memory caches, and object stores) offered by hybrid cloud providers in order to minimize the cost of data services~\cite{racs_socc10, spanstore_sosp13, scalia_sc12, fcfs_eurosys12, costlo_nsdi15, cast_hpdc15, pricing_hotcloud15, mos_hpdc16}. 
{\proj} takes a new route---exploiting the pay-per-use pricing model of FaaS to achieve pay-per-access storage.

\noindent\textbf{Garbage Collection in Storage Systems.} \addVLDBcomment{LFS~\cite{Rosenblum1992} designs a scavenging GC-style segment cleaner to defragment log segments in the file system. 
RAMCloud~\cite{ramcloud_fast14} inherits the idea 
and uses the cleaner to free the space in its log-structured memory.
Jiffy~\cite{jiffy} uses leases to identify live data and reclaims memory if the leases expire. Data repartitioning is needed for Jiffy to be elastic.
{\proj} is the first cloud memory storage that combines the traditional garbage collection technique and the FaaS properties to achieve fine-grained elasticity in disaggregated memory management.} 
\vspace{-6pt}
\section{Conclusion}
\label{sec:conclusion}

In this paper, we rethink the fundamental design of cloud storage systems and propose ServerlessMemory, a new cloud storage service. ServerlessMemory fills a gap in cloud storage services by offering a truly elastic, highly performant, yet low-cost with a pay-per-access memory storage layer.
By seamlessly combining ServerlessMemory with a cost-effective cloud object store, we present a generic and novel cloud storage system, {\proj}, which supports uninterrupted data services by employing an automated, fast, and reliable parallel recovery scheme.
Extensive evaluation via real-world production storage workloads and microbenchmarks
demonstrates that {\proj} outperforms existing systems while lowering costs with its pay-per-access pricing model.
Furthermore, {\proj} presents a new performance-\$cost tradeoff in the cloud storage landscape. 
{\proj} is open source and publicly available at: \url{https://github.com/ds2-lab/infinistore}.

\vspace{-6pt}
\begin{acks}
\label{sec:acknowledgements}
\end{acks}

We are grateful to the anonymous reviewers for their valuable feedback and comments. This work was sponsored in part by NSF grants: CNS-2045680 (an NSF CAREER Award), CAREER-2048044, an NSF CloudBank grant, CCF-1919075, CCF-1919113, OAC-2106446, and supported by an Adobe Research gift.

\label{startofrefs}
\clearpage
\newpage

{
\bibliographystyle{ACM-Reference-Format}

}

\label{startofappendix}
\newpage

\appendix
\section{Versioning}
\label{appendix:versioning}

\begin{figure}[h]
\vspace{-8pt}
\begin{center}
\begin{minipage}[b]{0.23\textwidth}
\begin{minted}[fontsize=\footnotesize, xleftmargin=2pt, numbersep=2pt,linenos]{go}
func boolean PUT(ckey, s):
 c := MT.Prepare(ckey,s,1)
 m := null
 for:
  m, ok := MT.Cas(ckey,m,c)
  if !ok and !m.IsDone():
   return m.Wait()
  elseif !ok:
   c.Revise(m.Ver+1)
   continue
  v := m.Ver
  MT.Store(ckey|v, m)
  sp := PB.Create(ckey|v, s)
  ok := SMS.Put(ckey|v, sp)
  if !ok: return m.Done(ok)
  async:
   do ok := COS.Put(ckey|v, sp)
   while !ok
   sp.Release()
  return m.Done(ok)
\end{minted}
\vspace{-12pt}
\end{minipage}
\begin{minipage}[b]{0.23\textwidth}
\begin{minted}[fontsize=\footnotesize, xleftmargin=2pt, numbersep=2pt,linenos]{go}
func stream GET(ckey):
 key := ckey
 for:
  m := MT.Load(key)
  if m == null: return null
  elseif !m.IsDoneOk():
   key := ckey|m.PrevVer()
   continue
  sp := PB.Load(ckey|m.Ver)
  if sp != null: return sp
  s := SMS.Get(ckey|v)
  if s != null: return s
  do val := COS.Get(ckey|v)
  while val == null
  SMS.Restore(ckey|v, val)
  return stream(val)
\end{minted}
\vspace{20pt}
\end{minipage}
\caption{The versioning algorithm for {\kvput}s and {\kvget}s with persistent buffer.
}
\label{fig:versioning}
\end{center}
\vspace{-10pt}
\end{figure}

{\proj} uses versioning to support updates and strong consistency.
The versioning algorithm with persistent buffer is presented in Figure~\ref{fig:versioning}.
For a {\kvput} request, the algorithm starts by preparing a candidate metadata for this object with version 1 (line 2), and then atomically inserts the candidate metadata into the daemon's metadata table (MT) using compare-and-swap (CAS) at line 5. If the insertion fails, current metadata is returned and the status of the metadata is checked. If a concurrent {\kvput} is executing (line 6), the algorithm waits till the end of the concurrent operation and returns a {\small\texttt{retry}} error so that the application may choose to retry this failed {\kvput} (line 7). 
If a new version is available (line 8), the algorithm revises the candidate metadata with a new version (line 9) and retry the metadata insertion at line 5. If the insertion succeeds, the algorithm then adds a newly versioned entry into the metadata table (line 12), creates a persistent buffer (PB) to intercept the input stream(s) (line 13), stores the data in SMS (line 14), and concludes the {\kvput} request (line 20). The algorithm also spawns a thread to asynchronously check if the data has been successfully persisted in COS (line 17-18); if yes, the thread will then release the persistent buffer (line 19).

For a {\kvget} request, the algorithm first loads the chunk metadata from the MT (line 4). If there is a concurrent {\kvput} for the same chunk that has not returned (line 6), the algorithm will retry to read the last successfully returned version (Line 7,4). The algorithm then checks if there is a hit in the persistent buffer for the corresponding chunk with the specified version (line 9).
If yes, it means that the chunk has not been persisted to COS (eventually it will be persisted), and the request would be served using the chunk buffered in the persistent buffer; this way, the read-after-write consistency is guaranteed (line 10). Otherwise, the chunk has already been persisted to COS; the request will be served from SMS. In rare cases the chunk needs to be restored from COS (line 11-16). In that case, the client keeps reading the chunk from the COS until a consistent chunk with the matching version is available (line 13-14). This looping-based technique is inspired by the consistency increasing algorithm proposed in SCFS~\cite{Bessani2014} in order to achieve strong consistency atop an eventually consistent COS. 

\end{document}